\documentclass{aa}
\usepackage{natbib}
\usepackage{graphics}
\usepackage{amssymb}
\usepackage{txfonts}
\begin{document}
\title{H$_2$CO and CH$_3$OH abundances in the envelopes around low-mass protostars}
\author{J.K. J{\o}rgensen\inst{1}\thanks{Present address:
Harvard-Smithsonian Center for Astrophysics, 60 Garden Street MS42,
Cambridge, MA 02138, USA } \and F.L. Sch\"{o}ier\inst{2} \and E.F. van
Dishoeck\inst{1}}

\institute{Leiden Observatory, P.O. Box 9513, NL-2300 RA Leiden, The
Netherlands \and Stockholm Observatory, AlbaNova, SE-106 91 Stockholm, Sweden}

\offprints{Jes K.\,J{\o}rgensen}
\mail{jjorgensen@cfa.harvard.edu} 
\date{Received <date> / Accepted <date>}

\abstract{ This paper presents the third in a series of single-dish
  studies of molecular abundances in the envelopes around a large
  sample of 18 low-mass pre- and protostellar objects. It focuses on
  typical grain mantle products and organic molecules, including
  H$_2$CO, CH$_3$OH and CH$_3$CN. With a few exceptions, all H$_2$CO
  lines can be fit by constant abundances of 7$\times
  10^{-11}$--8$\times 10^{-9}$ throughout the envelopes if ortho- and
  para lines are considered independently. The current observational
  dataset does not require a large H$_2$CO abundance enhancement in
  the inner warm regions, but this can also not be ruled out. Through
  comparison of the H$_2$CO abundances of the entire sample, the
  H$_2$CO ortho-para ratio is constrained to be 1.6$\pm$0.3 consistent
  with thermalization on grains at temperatures of 10--15~K. The
  H$_2$CO abundances can be related to the empirical chemical network
  established on the basis of our previously reported survey of other
  species and is found to be closely correlated with that of the
  nitrogen-bearing molecules. These correlations reflect the
  freeze-out of molecules at low temperatures and high densities, with
  the constant H$_2$CO abundance being a measure of the size of the
  freeze-out zone. An improved fit to the data is obtained with a
  `drop' abundance structure in which the abundance is typically a
  few~$\times$$10^{-10}$ when the temperature is lower than the
  evaporation temperature and the density high enough so that the
  timescale for depletion is less than the lifetime of the core. The
  location of the freeze-out zone is constrained from CO
  observations. Outside the freeze-out zone, the H$_2$CO abundance is
  typically a few~$\times 10^{-9}-10^{-8}$. The observations show that
  the CH$_3$OH lines are significantly broader than the H$_2$CO lines,
  indicating that they probe kinematically distinct regions. CH$_3$OH
  is moreover only detected toward a handful of sources and CH$_3$CN
  toward only one, NGC~1333-IRAS2. For NGC~1333-IRAS2, CH$_3$OH and
  CH$_3$CN abundance enhancements of two-three orders of magnitude at
  temperatures higher than 90~K are derived. In contrast, the
  NGC~1333-IRAS4A and IRAS4B CH$_3$OH data are fitted with a constant
  abundance and an abundance enhancement at a lower temperature of
  30~K, respectively. This is consistent with a scenario where
  CH$_3$OH probes the action of compact outflows on the envelopes,
  which is further supported by comparison to high frequency, high
  excitation CS $J=$~10--9 and HDO line profiles which uniquely probe
  warm, dense gas. The extent to which the outflow dominates the
  abundance enhancements compared with the passively heated inner
  envelope depends on the filling factors of the two components in the
  observing beam.

} \maketitle

\section{Introduction}
The chemistry of organic molecules in the envelopes around low-mass
protostars is likely to reflect directly in the molecular composition
of their circumstellar disks and eventual protoplanetary systems. A
number of competing mechanisms are important in regulating the
chemistry in these early deeply embedded stages: the heating of
protostellar cores due to central, newly formed stars results in
evaporation of ices in the innermost regions whereas shocks related to
the ubiquitous outflows may liberate ice mantles and trigger similar
effects but on larger scales. These mechanisms have been suggested to
be the cause of enhancements of, e.g., H$_2$O, H$_2$CO and CH$_3$OH on
small scales in the envelopes
\citep[e.g.,][]{ceccarelli98,ceccarelli00a,schoeier02,hotcorepaper}. This
paper, the third in a series, presents an analysis of, in particular,
H$_2$CO and CH$_3$OH abundances in the sample of 18 protostars studied
in a wide range of other molecules by
\cite{jorgensen02,paperii}. Those papers discussed observations of
molecular species predominantly probing the outer cold envelopes
around these objects. It was found that freeze-out at low temperatures
and high densities dominates the chemistry and that, in particular,
the freeze-out of CO is reflected in the abundances of a number of
related species at large distances from the central protostar. This
paper complements the study of H$_2$CO and CH$_3$OH in a subset of
objects by \cite{maret04,maret04ch3oh}. Both species are typical
grain-mantle products observed in interstellar ices. To fully
appreciate their chemistry it is important to compare their abundances
with the more general chemical network. For example, high resolution
observations of \cite{hotcorepaper} indicate that the H$_2$CO
abundance structures may be related to the ``drop abundance''
structures inferred from CO observations \citep{coevollet}. In these
drop abundance profiles, freeze-out occurs in a limited region of the
envelope where the temperature is low enough to prevent immediate
desorption (typically $<$~40 K) but the density still high enough that
the timescale for depletion is less than the lifetime of the core (see
also Fig.~10 of \cite{hotcorepaper}).

The class 0 protostar, IRAS~16293-2422, has long been the template for
astrochemistry studies of deeply embedded low-mass protostars due to
its rich spectrum \citep{blake94,vandishoeck95}. IRAS~16293-2422 has a
central warm and dense gas core where ices evaporate. Recently,
\cite{cazaux03} have shown the existence of a large number of complex
organic species in IRAS~16293-2422, further underscoring the rich
chemistry of this particular source. It remains an interesting
question whether this simply reflect ``first generation'' evaporation
of organic molecules at high temperatures or whether the timescales
are indeed long enough that a ``second generation'' hot core chemistry
can evolve in the innermost regions of these envelopes \citep[see,
e.g., discussion in][]{schoeier02}.

To put the IRAS~16293-2422 results in context, it is important to
expand the sample of well-studied protostars. \cite{paperii} presented
a survey of molecular species probing the cold outer component of
protostellar objects with different envelope masses, i.e., both
``class 0'' and ``class I'' objects. Although large variations in
abundances occur within the sample, it was found that IRAS~16293-2422
is in no way unique. The same objects were observed in transitions of
H$_2$CO and CH$_3$OH at the JCMT. \cite{maret04} presented the H$_2$CO
observations, together with observations from the IRAM~30~m telescope,
for a subset of exclusively class 0 objects. They reported the
existence (or possibility) of H$_2$CO abundance jumps (i.e., abundance
enhancements in the innermost regions of the envelopes at $T>90$~K
where all ices evaporate), in some cases up to four orders of
magnitude. However, \cite{hotcorepaper} showed through high angular
resolution data of IRAS~16293-2422 and L1448-C, that the exact
abundance structure of the outer envelopes may severely affect the
interpretation of the innermost envelope. For these low luminosity
sources the warm inner regions have diameters $< 100$~AU ($<
0.5$\arcsec), i.e., are significantly diluted for single-dish
observations with typical beam sizes of $15-30''$.

\cite{buckle02} studied the low excitation ($3_K-2_K$) lines of
CH$_3$OH toward a large sample of class 0 and I objects. They found
that a large fraction of the sources, predominantly the class 0
objects, show lines with two velocity components with CH$_3$OH being
enhanced by up to two orders of magnitude. \citeauthor{buckle02}
suggested that the broad component is due to outflow generated shocks
heating the envelope material and thus liberating the grain
mantles. Similar effects are also observed in several well-studied
``isolated'' outflows well separated from the protostars themselves
\citep{bachiller95,bachiller97,i2art}. These and other studies
illustrate that for CH$_3$OH, a big issue may be whether the abundance
enhancements derived from single-dish observations toward protostellar
cores are related to passive heating or the action of outflows.

This paper expands the work of \cite{maret04,maret04ch3oh} through
observations and modeling of H$_2$CO, CH$_3$OH and CH$_3$CN emission
for the entire sample of pre- and protostellar cores studied by
\cite{jorgensen02,paperii}, adopting the same physical models and
approach as in these papers. In addition, we present high excitation
CS $J=10-9$ and HDO observations which uniquely probe the dense and
warm gas in the envelope. Sect.~\ref{observations} presents the
observations forming the basis of this study. Sect.~\ref{models}
describes the model approach, highlighting the similarities and
differences with the work of \cite{maret04}. Sect.~\ref{discussion}
discusses the results, focusing on the distinction between passively
heated and shock processed material. Based on this study, it is
discussed which objects are good candidates for further studies of
low-mass protostars with a ``hot core'' chemistry.

\section{Observations}\label{observations}
\subsection{General issues}
The sample of 18 pre- and protostars presented by \cite{jorgensen02}
was observed in a wide range of lines at the James Clerk Maxwell
Telescope from 2001 through 2003. The observations of other species
besides H$_2$CO, CH$_3$OH and CH$_3$CN are discussed by
\cite{paperii}. \cite{maret04,maret04ch3oh} reported H$_2$CO and
CH$_3$OH observations, respectively, toward the sources with the most
massive envelopes - and hence strongest lines - in the sample, which
are also included in this paper. The following sections discuss the
observations of each of these species in detail. All lines were
observed and reduced in a standard way: pointing was checked regularly
at the telescope and typically found to be accurate to within a few
arcseconds. Calibration was checked by observations of line standards
and found to be accurate to within 20\%. The A3 and B3 receivers at
1.3 and 0.8~mm were used: the telescope beam sizes are typically
21\arcsec\ and 14\arcsec\ at these frequencies. The velocity
resolution ranged from 0.13 to 0.55~km~s$^{-1}$ for the different line
settings. Low order polynomials were subtracted and the spectra were
brought onto the $T_{\rm MB}$ scale through division by the main beam
efficiencies, $\eta_{\rm MB}$. Values of $\eta_{\rm MB}$ of 0.69
and 0.63 were adopted for the 230~GHz observations with receiver A3
and the 345~GHz observations with receiver B3, respectively.

Most of the observed lines are remarkably symmetric and for these
lines, a single Gaussian could be fitted. The only exceptions are a
few of the lowest excitation lines, which are integrated over $\pm
2$~km~s$^{-1}$ from the systemic velocity. This velocity interval
covers the emission from the quiescent envelope material as judged
from the optically thin Gaussian lines. The line intensities for all
species are given in Tables~\ref{h2co_int}--\ref{ch3cn_ints}. The
corresponding line widths are given in
Tables~\ref{h2co_widths}--\ref{ch3oh_widths} in the appendix. For the
non-detections, 3$\sigma$ upper limits are reported with
$\sigma=1.2\sqrt{\delta v\, \Delta_0 v}\,T_{\rm RMS}$ where $\delta v$
is the velocity resolution, $\Delta_0 v$ the expected line width to
zero intensity (assumed to be 4~km~s$^{-1}$), $T_{\rm RMS}$ the RMS
noise level for the given resolution and the factor 1.2 introducing
the 20\% calibration uncertainty.

In addition to these observations, the high frequency RxW receiver was
used to observe CS $J=10-9$ at 489.751~GHz for four sources (L1448-C,
NGC~1333-IRAS2, -IRAS4A and -IRAS4B) over two nights in November
2002. Special care was taken with the calibration: comparison with
nearby spectral standards was found to vary by $< 20$\% over these two
nights, during which the sky opacity was $\lesssim 0.05$ at 225~GHz
and the elevation of the sources higher than $\approx
50^\circ$. Still, high system temperatures of up to $\sim$~5000~K were
found and this, together with the pointing uncertainties of a few
arcseconds of the JCMT (compared to a beam size of 10\arcsec), may
cause the absolute calibration to be somewhat uncertain for these
observations. The CS 10--9 lines were detected toward
NGC~1333-IRAS4A and IRAS4B but not L1448-C and NGC~1333-IRAS2. The
implications of these observations are discussed further in
Sect.~\ref{cs_hdo}.

\subsection{H$_2$CO}
In addition to the observations presented by \cite{maret04}, H$_2$CO
emission from the ortho $5_{15}-4_{14}$ line at 351.768~GHz and the
para $5_{05}-4_{04}$ line at 362.736~GHz was observed for all
sources. Furthermore JCMT archival data exist for the para
$3_{03}-2_{02}$ and $3_{22}-2_{21}$ lines at 218.222 and 218.475~GHz
for most sources and these observations were included for those
sources not observed at the IRAM 30m by
\cite{maret04}. Table~\ref{h2co_int} lists the line intensities for
all sources. The typical line widths of the H$_2$CO lines are
1--2~km~s$^{-1}$ (FWHM). For the prestellar cores, L1544 and L1689B,
intensities of $J=2-1$ and $3-2$ lines from IRAM 30~m
observations by \cite{bacmann03} were included in the modeling.
\begin{table}
\caption{Integrated H$_2$CO line intensities ($\int T_{\rm MB}\, {\rm d}v$
  [K~km~s$^{-1}$]).}\label{h2co_int}
\begin{center}\begin{tabular}{llllllll} \hline\hline
 & \multicolumn{3}{c}{p-H$_2$CO} & \multicolumn{3}{c}{o-H$_2$CO} \\
       & $3_{03}-2_{02}$$^{b}$ & $3_{22}-2_{21}$$^{b}$ & $5_{05}-4_{04}$ & $5_{15}-4_{14}$ \\ \hline
L1448-I2         & $\ldots$ &$\ldots$&$<$0.9  &  1.2  \\
L1448-C$^{a}$    &  3.4      &  0.4   &  1.3   &  1.0  \\
N1333-I2$^{a}$   &  4.9      &  1.0   &  1.8   &  1.6  \\
N1333-I4A$^{a}$  &  9.3      &  2.2   &  2.9   &  5.5  \\
N1333-I4B$^{a}$  &  9.6      &  4.7   &  5.9   &  7.5  \\
L1527$^{a}$      &  3.0      &  0.2   &  0.4   &  1.0  \\
VLA1623$^{a}$    &  5.0      &$\ldots$&  1.2   &  0.9  \\
L483             &  2.3      & $<0.2$ &  1.2   &  1.3  \\
L723             &  1.1$^c$  & $<0.1$$^c$ & 1.1 &  2.0  \\
L1157$^{a}$      &  1.1      & $<0.3$ &  0.5   &  1.2  \\
CB244            &  0.9      & $<0.4$ &  0.6   &  1.4  \\
L1489            & $<$0.3    &$<$0.3  &$<$0.4  &  0.7  \\
TMR1             &  0.6      &$<$0.1  &$<$0.3  &  0.4  \\
L1544            & 0.3$^{d}$ & $\ldots$ &$<$0.5  &$<$0.4 \\
L1689B           & 1.7$^{d}$ & 1.0$^{d}$  &$<$0.3  &$<$0.4 \\ \hline
\end{tabular}\end{center}

$^{a}$Observations previously reported by \cite{maret04}. For some of
these sources additional o-H$_2$CO $2_{12}-1_{11}$ and
$4_{14}-3_{13}$, p-H$_2$CO $5_{24}-4_{23}$ and o-H$_2^{13}$CO
observations and limits were reported by
\cite{maret04}. $^{b}$$3_{03}-2_{02}$ and $3_{22}-2_{21}$ line
intensities from \cite{maret04} are from IRAM 30~m
observations. $^c$From SEST (HPBW=24\arcsec) observations. $^{d}$IRAM
30~m observations from \cite{bacmann03}: for these two pre-stellar
cores o-H$_2$CO $2_{12}-1_{11}$ observations were also reported by
\citeauthor{bacmann03} and included in the modeling.
\end{table}

\subsection{CH$_3$OH}
The CH$_3$OH $7_K-6_K$ observations of the sources in NGC~1333 are
presented in \cite{maret04ch3oh}. In addition, the $5_K-4_K$ band at
241~GHz was observed for the NGC~1333 sources. The line intensities
from the $5_K-4_K$ band data are given in
Table~\ref{ch3oh_int1333}. Table~\ref{ch3oh_intb} lists the CH$_3$OH
observations for the remaining sources. Only the NGC~1333 sources have
detections for a large number of CH$_3$OH lines. Besides these, only
L723, L1448-C and VLA~1623 show detections in the $7_K-6_K$ band and
only of 1--2 lines each. The HDO $2_{11}-2_{12}$ transition at
241.5616 GHz was covered in the CH$_3$OH $5_K-4_K$ setting, but was
only detected toward NGC~1333-IRAS2.

As for H$_2$CO, the CH$_3$OH lines can be divided into two forms, A-
and E-type, depending on the rotation of the methyl CH$_3$-group with
respect to the OH group. Some systematics, which may give clues to the
origin of the CH$_3$OH emission, can be directly inferred from these
results. For example, NGC~1333-IRAS2 shows the highest excitation
lines for both the A and E-type for the $5_K-4_K$ and $7_K-6_K$
transitions compared to NGC~1333-IRAS4A and -IRAS4B. Still, the lowest
excitation lines toward NGC~1333-IRAS2 are weaker than those toward
the NGC~1333-IRAS4 sources: barring pure excitation/opacity effects,
this could be interpreted as a warmer interior, possibly with higher
abundances in NGC~1333-IRAS2. In particular, since the sources
otherwise have relatively similar physical structures these
differences suggest that one or more sources will show chemical
gradients. Furthermore the CH$_3$OH lines are significantly broader
with line widths of 4--6~km~s$^{-1}$ (FWHM) compared to 1--2~km~s$^{-1}$ found
for H$_2$CO and the other species in \cite{jorgensen02, paperii}.
\begin{table}
\caption{CH$_3$OH $5_K-4_K$ line intensities ($\int T_{\rm MB}\,{\rm
d}v$ [K~km~s$^{-1}$]) for sources in NGC~1333.}\label{ch3oh_int1333}
\begin{center}\begin{tabular}{lllll} \hline\hline
Line & Frequency &             IRAS2  & IRAS4A & IRAS4B \\ \hline
\multicolumn{5}{c}{$5_K-4_K$ band; E-type} \\ \hline
    +0E &           241.7002 & 0.41   & 2.9    & 2.1    \\
  $-$1E & \phantom{241}.7672 & 0.86   & 7.8    & 4.6    \\
  $-$4E & \phantom{241}.8132 & 0.17   &$<$0.06 &$<$0.06 \\
    +4E & \phantom{241}.8296 & 0.16   &$<$0.06 &$<$0.06 \\
    +3E & \phantom{241}.8430 & 0.30   & 0.48   & 0.55   $\dagger$  \\
  $-$3E & \phantom{241}.8523 & 0.17   &$<$0.06 &$<$0.06 \\
    +1E & \phantom{241}.8790 & 0.33   & 1.3    & 1.3    \\
$\pm$2E & \phantom{241}.9044 & 0.45   & 2.7    & 2.3    \\ \hline
\multicolumn{5}{c}{$5_K-4_K$ band; A-type} \\ \hline
    +0A &           241.7914 & 1.1    & 8.9    & 5.3    \\
$\pm$4A & \phantom{241}.8065 & 0.18   &$<$0.06 &$<0.06$ \\
$\pm$3A & \phantom{241}.8329 & 0.33   & 0.64   & 0.53   \\
  $-$2A & \phantom{241}.8430 & 0.30   & 0.48   & 0.55   $\dagger$  \\
    +2A & \phantom{241}.8877 & 0.23   & 0.49   & 0.50   \\ \hline
\end{tabular}\end{center}

``$\dagger$''The $5-4$ +3E and $-$2A lines are blended at 241.8430 GHz
and have therefore (although observed) been excluded from the
modeling. The quoted intensity refers to the total intensity of both
lines.
\end{table}

\begin{table}
\caption{CH$_3$OH line intensities ($\int T_{\rm MB}\,{\rm
d}v$ [K~km~s$^{-1}$]) and limits for sources not in
  NGC~1333.}\label{ch3oh_intb}
\begin{center}\begin{tabular}{lll}\hline\hline
                 & \multicolumn{2}{c}{7--6}    \\ 
                 &  -1E  & 0A+   \\ \hline
L1448-I2         & \multicolumn{2}{c}{$<0.09$} \\
L1448-C          & 0.254 & 0.415 \\
L1527            & \multicolumn{2}{c}{$<0.12$} \\
VLA1623          & 0.063 & 0.060 \\
L483             & \multicolumn{2}{c}{$<0.18$} \\
L723             &$<0.06$& 0.373 \\
L1157            & \multicolumn{2}{c}{$<0.18$} \\
CB244            & \multicolumn{2}{c}{$<0.12$} \\
L1489            & \multicolumn{2}{c}{$<0.09$} \\
TMR1             & \multicolumn{2}{c}{$<0.09$} \\ \hline
\end{tabular}\end{center}
\end{table}

\subsection{CH$_3$CN}
\begin{figure}
\resizebox{\hsize}{!}{\includegraphics{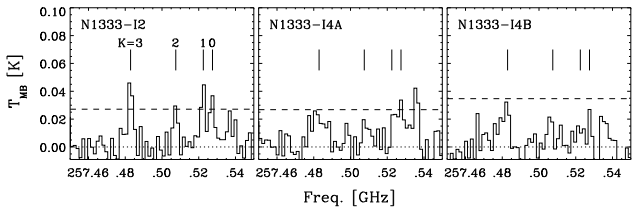}}
\caption{CH$_3$CN $14_K-13_K$ line observations at 257.5~GHz of the
NGC~1333 sources. The vertical lines indicate the expected locations
of the $K=0,1,2,3$ lines. The dashed line indicates the 3$\sigma$
detection limit.}\label{ch3cnobs}
\end{figure}

CH$_3$CN $14_K-13_K$ was observed at 257.5~GHz for the NGC~1333
sources with 2 hours integration time per source, reaching RMS levels
of 0.02~K ($T_{\rm MB}$) in 0.36~km~s$^{-1}$ channels. At these levels the
$K=0,\ldots,3$ components are detected for NGC~1333-IRAS2, as shown in
Fig.~\ref{ch3cnobs}. Since NGC~1333-IRAS2 typically shows weaker lines
for other molecules than the two NGC~1333-IRAS4 sources \citep[see,
  e.g.,][]{paperii}, this suggests that CH$_3$CN may probe a different
chemical regime (together with the CH$_3$OH lines) than the bulk, cold
envelope in this source. Recently \cite{bottinelli04n1333i4a} reported
detections at the 40-70~mK ($T_{\rm MB} {\rm (peak)}$) of the CH$_3$CN
$14_K-13_K$ components from IRAM~30~m observations toward
NGC~1333-IRAS4A. These are consistent with our upper limits taking
into account the smaller beam size of the IRAM~30~m telescope.

\begin{table}
\caption{CH$_3$CN and CH$_3$OCH$_3$ line intensities ($\int T_{\rm
MB}\,{\rm d}v$ [K~km~s$^{-1}$]) and 3$\sigma$ upper limits for the sources in
NGC~1333.}\label{ch3cn_ints}
\begin{center}\begin{tabular}{lllll} \hline\hline
Line & Frequency & IRAS2 & IRAS4A & IRAS4B \\ \hline
\multicolumn{5}{l}{CH$_3$CN} \\
14$_3$--13$_3$ & 257.4828 & 0.20     & $<$0.09$^{a}$  & $<$0.1    \\
14$_2$--13$_2$ & 257.5076 & 0.10     & --        & --        \\
14$_1$--13$_1$ & 257.5224 & 0.17     & --        & --        \\
14$_0$--13$_0$ & 257.5274 & 0.15     & --        & --        \\ \hline
\multicolumn{5}{l}{CH$_3$OCH$_3$} \\
13$_{1,13}$--12$_{0,12}$ & 241.9468 & $<$0.08  & $<$0.1    & $<$0.1    \\ \hline
\end{tabular}\end{center}

$^{a}$\cite{bottinelli04n1333i4a} report detection of the 14$_K$--13$_K$
lines at the 40-70~mK level from IRAM~30m observations toward
NGC~1333-IRAS4A.
\end{table}

\section{Modeling}\label{models}
To model the chemical abundances the approach described in
\cite{jorgensen02,paperii} and \cite{schoeier02}, and utilized for the
entire sample of sources and molecules, was adopted. Each species was
modeled with the envelope physical structure from \cite{jorgensen02}
derived from dust radiative transfer modeling of their submillimeter
(SCUBA) continuum emission and SEDs. The line radiative transfer was
then performed using the code of \cite{schoeier02} constraining the
average molecular abundances. This code was benchmarked to high
accuracy against a large number of other line radiative transfer codes
for a number of test problems by \cite{vanzadelhoff02} and found to
agree within the Monte Carlo noise.

A number of different trial abundance profiles are considered
besides profiles with a constant abundance, $X_0$, of a given molecule
at all radii (see also Fig.~10 of \cite{hotcorepaper}). In ``jump''
profiles the abundance increase from $X_0$ to $X_{\rm J}$ at a given
temperature in the envelope, e.g., simulating ice mantle
evaporation. Likewise ``anti-jump'' profiles have a decrease in
abundance in the inner region. Finally, in ``drop'' profiles the
abundance $X_0$ is high (comparable to the undepleted abundance
expected in a warm molecular cloud without significant freeze-out), in
the inner and outermost regions of the envelope. At radii where the
temperature is lower than a molecule-specific desorption temperature
$T_{\rm ev}$ and the density higher than the density $n_{\rm de}$
where the timescale for freeze-out is lower than the lifetime of the
core, the abundance decreases to $X_{\rm D}$.

For H$_2$CO the collisional rate coefficients used in
\cite{schoeier02} were adopted. As described by \cite{maret04}, ortho
H$_2^{13}$CO lines are detected for the three sources in NGC~1333. The
corresponding abundances were likewise calculated and taken into
account in subsequent discussions. For CH$_3$OH, new collisional rate
coefficients by \cite{pottage04} were used. For CH$_3$CN LTE
excitation was assumed. The molecular data files are taken from the
\emph{Leiden Atomic and Molecular Database} \citep{schoeier03radex}
and are publically
available\footnote{\tt{http://www.strw.leidenuniv.nl/$\sim$moldata}}. Each
of the ortho- and para H$_2$CO and the A- and E-type CH$_3$OH are
treated as separate molecules, which is possible since radiative
transitions between the different species are ruled out. In the first
iteration, the abundances are kept constant throughout the envelope
and the results from the best fits are given in
Tables~\ref{abund_h2co}-\ref{abund_ch3cn} below.
\begin{table*}
\caption{Inferred abundances for o-H$_2$CO, p-H$_2$CO and o-H$_2^{13}$CO, reduced
$\chi^2$ and number of lines where applicable.}\label{abund_h2co}
\begin{center}
\begin{tabular}{llllllll}\hline\hline
Source  & $M_{\rm env}$ & \multicolumn{3}{c}{o-H$_2$CO} & \multicolumn{3}{c}{p-H$_2$CO}  \\ 
 &  $M_\odot $ & Abundance       & $\chi^2_{\rm red}$ & $n_{\rm lines}$ & Abundance       & $\chi^2_{\rm red}$ & $n_{\rm lines}$ \\ \hline
\multicolumn{7}{c}{Class 0 objects ($M_{\rm env} > 0.5 M_\odot$)} \\
L1448-I2  & 1.5  & 3.0$\times 10^{-10}$    & $\ldots$ &  1     & $<$1.8$\times 10^{-10}$  &$\ldots$ &  (1)  \\
L1448-C   & 0.93 & 8.9$\times 10^{-10}$    & 1.2      &  3     &  6.8$\times 10^{-10}$  & 1.0     &  4    \\
N1333-I2  & 1.7  & 4.3$\times 10^{-10}$    & 0.63     &  3     &  3.9$\times 10^{-10}$  & 1.3     &  4    \\
N1333-I4A & 2.3  & 3.4$\times 10^{-10}$    & 0.59     &  3     &  2.3$\times 10^{-10}$  & 3.1     &  4    \\
N1333-I4B & 2.0  & 8.7$\times 10^{-10}$    & 6.2      &  3     &  4.1$\times 10^{-9}$   & 0.96    &  4    \\
L1527     & 0.91 & 8.4$\times 10^{-10}$    & 7.7      &  3     &  8.4$\times 10^{-10}$  & 0.44    &  3    \\
VLA1623   & 0.22 & 7.7$\times 10^{-10}$    & 0.34     &  3     &  1.3$\times 10^{-9}$   & 0.061   &  2    \\
L483      & 4.4  & 6.3$\times 10^{-10}$    & $\ldots$ &  1     &  3.3$\times 10^{-10}$  & 6.7     &  2    \\
L723      & 0.62 & 3.1$\times 10^{-9}$     & $\ldots$ &  1     &  9.6$\times 10^{-10}$  & 2.6     &  2    \\
L1157     & 1.6  & 9.2$\times 10^{-11}$    & 3.9      &  3     &  3.6$\times 10^{-11}$  & 2.1     &  2    \\[1.5ex]
\multicolumn{7}{c}{Class I objects ($M_{\rm env} < 0.5 M_\odot$)} \\[1.ex]
CB244     & 0.28 & 5.3$\times 10^{-9}$     & $\ldots$ &  1     &  6.9$\times 10^{-10}$  & 10.9    &  2    \\
L1489     & 0.097& 2.7$\times 10^{-9}$     & $\ldots$ &  1     &$<$1.4$\times 10^{-9}$  & $\ldots$&  (1)  \\
TMR1      & 0.12 & 5.0$\times 10^{-9}$     & $\ldots$ &  1     &  3.4$\times 10^{-9}$   & $\ldots$&  1+(1)\\[1.5ex]
\multicolumn{7}{c}{Pre-stellar cores} \\[1.5ex]
L1544$^a$ & 2.5  & 4.0$\times 10^{-11}$    & $\ldots$ &  1     &  3.0$\times 10^{-11}$  & $\ldots$&  1    \\ 
L1689B$^a$& 2.5  & 2.0$\times 10^{-10}$    & 8.4      & 2+(1)  &  1.2$\times 10^{-10}$  & 0.6     &  2    \\ \hline
          & \multicolumn{3}{c}{o-H$_2^{13}$CO}  & \multicolumn{3}{c}{} \\ \hline
N1333-I2  & 1.7  & 3.0$\times 10^{-11}$    & 3.3      &  2     & \multicolumn{3}{c}{} \\
N1333-I4A & 2.3  & 5.2$\times 10^{-12}$    & 2.5      &  3     & \multicolumn{3}{c}{} \\
N1333-I4B & 2.0  & 8.0$\times 10^{-11}$    & $\ldots$ &  1     & \multicolumn{3}{c}{} \\ \hline
\end{tabular}
\end{center}

$^{a}$Quoted value based on line intensities reported by
\cite{bacmann03}. Upper limits from JCMT lines only are one to two
orders of magnitude larger.
\end{table*}
\begin{table*}
\caption{Inferred abundances for CH$_3$OH, reduced $\chi^2$ and
number of lines where applicable.}\label{abund_ch3oh}
\begin{center}
\begin{tabular}{llllllll}\hline\hline
Source    & \multicolumn{3}{c}{CH$_3$OH A-type} & \multicolumn{3}{c}{CH$_3$OH E-type} \\
          &   Abundance     & $\chi^2_{\rm red}$ & $n_{\rm lines}$ &   Abundance     & $\chi^2_{\rm red}$ & $n_{\rm lines}$ \\ \hline
\multicolumn{7}{c}{Class 0 objects ($M_{\rm env} > 0.5 M_\odot$)} \\[1.5ex]
L1448-I2  & $<$1.1$\times 10^{-10}$ & $\ldots$     &  (1)  &$<$1.4$\times 10^{-10}$ & $\ldots$ &  (1)          \\
L1448-C   &  1.3$\times 10^{-9}$$^a$& $\ldots$     &   1   & 1.0$\times 10^{-9}$$^{a}$    & $\ldots$ &   1     \\
N1333-I2  &  1.4$\times 10^{-9}$$^a$&  20.4        &   8   & 1.3$\times 10^{-9}$$^{a}$    & 18.4     &  13     \\
N1333-I4A &  2.9$\times 10^{-9}$$^a$&   4.5        &   7   & 2.5$\times 10^{-9}$$^{a}$    & 3.1      &   7     \\
N1333-I4B &  3.5$\times 10^{-8}$$^a$&  11.4        &   7   & 9.1$\times 10^{-9}$$^{a}$    & 9.6      &   7     \\
L1527     & $<$5.5$\times 10^{-10}$ & $\ldots$     &  (1)  &$<$6.8$\times 10^{-10}$ & $\ldots$ &  (1)          \\
VLA1623   &  2.5$\times 10^{-10}$   & $\ldots$     &   1   & 3.4$\times 10^{-10}$   & $\ldots$ &   1           \\
L483      & $<$2.2$\times 10^{-10}$ & $\ldots$     &  (1)  &$<$2.7$\times 10^{-10}$ & $\ldots$ &  (1)          \\
L723      & 1.8$\times 10^{-9}$     & $\ldots$     &   1   &$<$3.6$\times 10^{-10}$ & $\ldots$ &  (1)          \\
L1157     & $<$2.7$\times 10^{-10}$$^a$ & $\ldots$ &  (1)  &$<$3.4$\times 10^{-10}$$^{a}$ & $\ldots$ &  (1)    \\[1.5ex]
\multicolumn{7}{c}{Class I objects ($M_{\rm env} < 0.5 M_\odot$)} \\[1.5ex]
CB244     & $<$1.0$\times 10^{-9}$  & $\ldots$     &  (1)  &$<$1.3$\times 10^{-9}$  & $\ldots$ &  (1)          \\
L1489     & $<$9.9$\times 10^{-10}$ & $\ldots$     &  (1)  &$<$1.2$\times 10^{-9}$  & $\ldots$ &  (1)          \\
TMR1      & $<$2.2$\times 10^{-9}$  & $\ldots$     &  (1)  &$<$2.8$\times 10^{-9}$  & $\ldots$ &  (1)          \\ \hline
\end{tabular}
\end{center}

$^{a}$See also \cite{maret04ch3oh}.
\end{table*}
\begin{table}
\caption{Inferred abundances and reduced $\chi^2$ from models of
CH$_3$CN toward IRAS2.}\label{abund_ch3cn}\label{abund_ch3och3}
\begin{center}\begin{tabular}{llll}\hline\hline
Source    & Abundance          & $\chi^2_{\rm red}$ & $n_{\rm lines}$ \\ \hline
\multicolumn{4}{c}{CH$_3$CN} \\
N1333-I2  &\phantom{$<$}8.3$\times 10^{-11}$       & 8.5      &   4     \\
          &\phantom{$<$}7$\times 10^{-9}$$^a$      & 1.4      &   4     \\ \hline
\end{tabular}\end{center}

$^a$Abundance in inner ($T > 90$~K) region. 3$\sigma$ upper
limit on the abundance in the outer envelope of 2$\times 10^{-11}$. 
\end{table}

\subsection{H$_2$CO}\label{h2co_model}
For most sources the H$_2$CO line intensities are well-fit with
constant abundance models for each of the p-H$_2$CO and o-H$_2$CO
species. For a few sources, one of these two has a $\chi^2_{\rm red}$
higher than 3 (o-H$_2$CO: NGC~1333-IRAS4B, L1527, L1157 and L1689B;
p-H$_2$CO: NGC~1333-IRAS4A, L483 and CB244). This is in contrast to
\cite{maret04} who inferred large abundance jumps for the studied
sources. As shown below this is due to a number of differences in the
assumptions between the work of \cite{maret04} and this study, in
particular the ortho-para ratio. Below we discuss some of these
differences.

\subsubsection{Velocity structure} \cite{maret04} assume a
non-turbulent but infalling envelope, whereas our work assumes a
constant turbulent broadening of 0.5--1~km~s$^{-1}$ throughout the
envelopes reproducing the observed line widths. The derived abundances
do not depend on the velocity field if only optically thin lines are
considered \citep{paperii}. For the constant abundances listed above,
the observed lines have typical optical depths of 0.1--1. This
constant width may not be an adequate description of emission
originating from the innermost dense regions in the case of large
abundance jumps, but in the context of an inside-out collapsing
envelope, the outer envelope will naturally be characterized by the
turbulent broadening. In class 0 objects typical inferred ages are
$\sim$~10$^4$ years, which for typical values of the sound speed of
0.3--0.5~km~s$^{-1}$ translate to infall radii of 500--1000~AU. Such
sizes are unresolved by single-dish observations and even by
medium-high resolution interferometer data. This implies that the
infall radius encompasses the hot inner region, but that most of the
mass in the envelope material is not infalling.

\begin{figure}
\resizebox{\hsize}{!}{\includegraphics{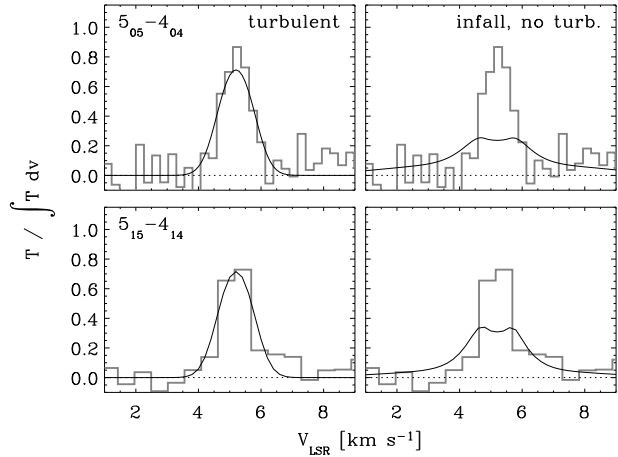}}
\caption{Spectra and modeled line profiles of the $5_{05}-4_{04}$
(para) and $5_{15}-4_{14}$ (ortho) H$_2$CO lines (upper and lower
panels, respectively) toward L1448-C in a pure turbulent, constant
abundance envelope (this paper; left) and in a non-turbulent envelope
with an abundance jump infalling toward a $0.5$~$M_\odot$ central
object (\cite{maret04}; right). The spectra have all been normalized
by division with the total integrated line intensity to bring out more
clearly the comparison between the actual line
shapes.}\label{h2colineprofile}
\end{figure}

From line radiative transfer it is possible to make exact predictions
for the line profiles, which can be compared to the observed spectra
to constrain the velocity field. Fig.~\ref{h2colineprofile} compares
the observed $5_{05}-4_{04}$ and $5_{15}-4_{14}$ spectra toward
L1448-C with the modeled line profiles adopting the non-infalling but
turbulent envelope model (with constant H$_2$CO abundance) from this
paper and the non-turbulent, infalling model (with abundance jump)
from \cite{maret04}. Considering only the higher excitation H$_2$CO
$J=5-4$ lines should limit confusion from the surrounding material. It
is seen that the ``turbulent'' envelope model clearly provides the
closest match to the observed line profile, whereas the non-turbulent,
infalling model provides too much broadening. It is interesting to
note that the same turbulent broadening which is used to fit
low-excitation CO isotopic lines from the cold, outer envelope
\citep{jorgensen02} works well in also explaining the observed H$_2$CO
line widths. In the context of, e.g., an inside-out collapsing
envelope as in \cite{shu77} this suggests that the infalling region of
the envelope is small and that even the relatively high excitation
H$_2$CO lines studied in this paper are still predominantly sensitive
to the outer envelope. It is worth re-emphasizing that both turbulent
and non-turbulent/infalling models give a good constant abundance fit
when the ortho and para lines are treated independently and that
abundance jumps are in general not required for the studied sources,
other than in the context of a ``drop model'' (see
Sect.~\ref{othermolecules}). Naturally a combination of turbulent
broadening and systematic motions (e.g., infall) most likely applies
to the studied envelopes and a fully self-consistent model explaining
also emission of optically thick lines would have to take these into
account. Observations of optically thick transitions of, e.g., HCO$^+$
could be used to constrain such models, but this is beyond the scope
of the present study.

\subsubsection{H$_2$CO ortho-para ratio}\label{orthopararatio}
\begin{figure}
\resizebox{\hsize}{!}{\includegraphics{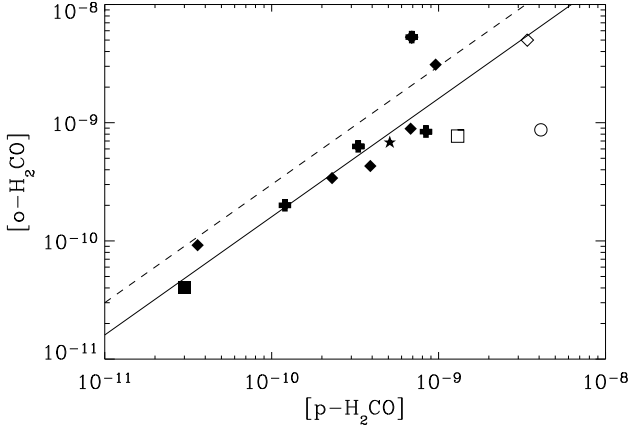}}
\caption{Comparison between the abundances of the p-H$_2$CO and
  o-H$_2$CO species. The class 0 objects are indicated by filled
  diamonds, the class I objects by open diamonds and the pre-stellar
  cores by filled squares. The class 0 objects \object{VLA1623} and
  \object{IRAS~16293-2422} have been singled out by an open square and
  star, respectively. Sources with poor fits ($\chi^2_{\rm red} \ge
  5$) to either the p-H$_2$CO or o-H$_2$CO species are indicated by
  black crosses. Of these N1333-I4B has further been singled-out with
  the open circle. The statistical uncertainties from the fits
  including calibration uncertainties are 10--20\%, i.e., the error
  bars are of similar size as the plot symbols. The solid line
  indicates the best fit linear correlation between the two sets of
  abundances (excluding the poorly fit sources), corresponding to an
  ortho-para ratio of 1.6:1. The dashed line indicates the relation
  for an ortho-para ratio of 3:1.}\label{orthopara_abund}
\end{figure}

A second difference with \cite{maret04} is the assumption of a fixed
ortho/para ratio. Since para and ortho H$_2$CO can be considered to be
separate molecules, their abundances can be determined
independently. Fig.~\ref{orthopara_abund} compares the abundances for
the two species. A very close correlation exists (Pearson correlation
coefficient of 0.9; see also Sect.~\ref{othermolecules}), which can be
fitted by an ortho-para ratio of 1.6$\pm 0.3$. The very tight
correlation indicates that both species probe the same region of the
envelope and that their abundance ratios are established under similar
conditions in all sources.

\begin{figure}
\resizebox{\hsize}{!}{\includegraphics{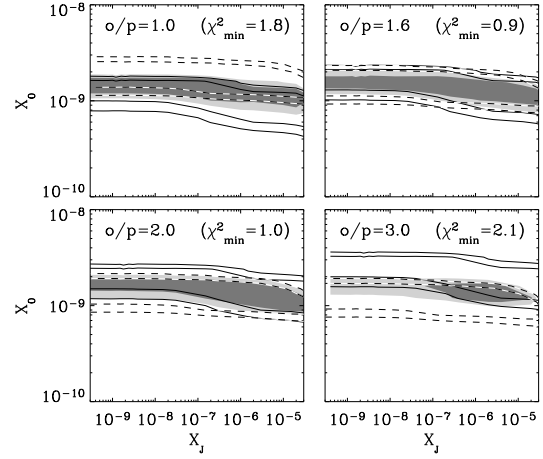}}
\caption{Constraints on the H$_2$CO abundances in the inner ($T>90$~K)
and outer ($T<90$~K) envelope ($X_J$ and $X_0$, respectively) and
effect of adopted ortho-para ratio for L1448-C. A non-turbulent,
free-falling envelope has been adopted as in \cite{maret04}. The solid
line contours indicate the 2 and 4$\sigma$ confidence levels for
ortho-H$_2$CO, whereas the dashed line contours indicate the
corresponding confidence levels for p-H$_2$CO. The grey scale contours
indicate similar confidence levels for the H$_2$CO abundance combining
the constraints from the two datasets and assuming the ortho-para
ratios of 1.0, 1.6, 2.0 and 3.0, respectively, as indicated in the top
of each panel. }\label{l1448c_orthoparaconfidence}
\end{figure}

\begin{figure}
\resizebox{\hsize}{!}{\includegraphics{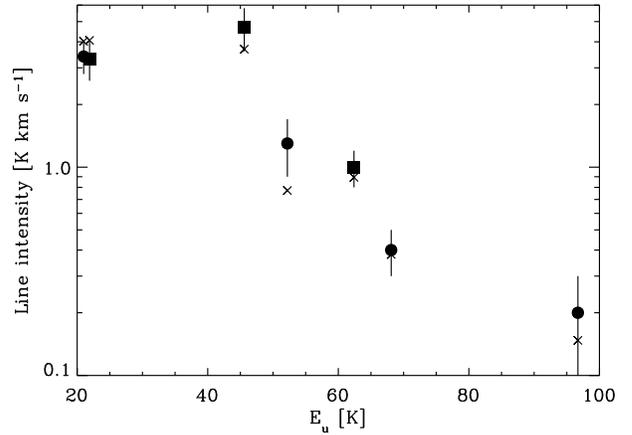}}
\caption{Comparison between observed and modeled line intensities
as function of the energy of the upper level for each H$_2$CO
transition toward L1448-C. The observed intensities of the para lines
are indicated by the circles, the intensities of ortho lines by the
squares and the modeled intensities by the crosses.}\label{ener_mod}
\end{figure}

\cite{maret04} assumed an ortho-para ratio of 3 to combine o-H$_2$CO
and p-H$_2$CO line observations to constrain the abundance
structure. Whereas this gives in principle fewer free parameters in
the modeling, one should be careful when interpreting the
results. This is clearly illustrated in
Fig.~\ref{l1448c_orthoparaconfidence} where constraints on the total
H$_2$CO abundance from ortho and para lines for L1448-C are shown. It
can be seen that the overlapping confidence levels depend critically
on the adopted ortho-para ratio, with a high ortho-para ratio driving
an increased abundance jump in the inner envelope up to 4 orders of
magnitude. For both species a constant abundance model provides a good
fit to the observed lines and the combination of the two suggests an
ortho-para ratio of 1.6, in agreement with the conclusion
above. Fig.~\ref{ener_mod} compares the modeled and observed line
intensities for L1448-C as functions of energy level of the observed
H$_2$CO transitions. No systematics are seen (for example, transitions
from higher levels are not systematically underestimated in the
models) and almost all observed intensities are well reproduced within
the observational uncertainty. In general, it is difficult to
  constrain the ortho-para ratio for a specific source because of the
  intrinsic uncertainty in the observations and modeling, including
  varying ortho-para ratios with position and varying optical depth
  \citep{schoeier02}. The strength of our analysis is that the ratio
  is based on a large sample of sources, statistically reducing some
  of these uncertainties.

The ortho-para ratio itself contains interesting information about the
H$_2$CO formation, as discussed by \cite{kahane84}. At high
temperatures the ortho-para ratio approaches the relative statistical
weights of 3:1, but thermalization at lower temperatures ($T \sim
10-15$~K) makes the two abundances closer to equal. As argued by
\citeauthor{kahane84}, gas-phase thermalization is improbable: the
life-times of the established ortho-para species are significantly
longer than, e.g., the time it takes to destroy a given H$_2$CO
molecule through reactions with molecular ions. \citeauthor{kahane84}
furthermore find that chemical reactions lead to ortho-para ratios of
3--5. In contrast, H$_2$CO can be thermalized on dust grains and
released subsequently. For grain temperatures of 10--15~K, an
ortho-para ratio of 1.5 is in good agreement with the results from
this paper. This is also what is found in typical dark clouds such as
TMC1 and L134N, whereas warmer regions such as the Orion clouds show
ortho-para ratios closer to the statistical 3:1 ratio
\citep{kahane84,mangum93}. The derived ortho-para ratio of 1.6$\pm$0.3
for the sources in this sample therefore suggests that the observed
H$_2$CO lines are predominantly sensitive to the outer cold envelope
and that relative abundances of the ortho and para species are
established there.

\subsubsection{Comparison to other molecules and implications for abundance structures}\label{othermolecules}
To quantify the relations between the abundances of the observed
molecular species, \cite{paperii} calculated Pearson correlation
coefficients for each pair of abundances. The Pearson correlation
coefficient, $P$, is a measure of how well a $(x,y)$ data set is fit
by a linear correlation compared to the spread of $(x,y)$
points. Values of $\pm 1$ indicate good correlations (with positive or
negative slopes) whereas a value of 0 indicates no correlation. In our
studies of other molecules, strong correlations ($|P| \ge 0.7$) were
found between molecules for which relations were expected based on
chemical considerations, for example between CO and HCO$^+$ or between
the sulfur-bearing species. To extend this discussion, correlation
coefficients were calculated between the abundances found in this
paper and those from \cite{paperii} (see Table~\ref{pearson}).

\begin{table}
\caption{Pearson correlation coefficients between abundances found in
this paper and abundances from \cite{paperii}.}\label{pearson}
\begin{center}\begin{tabular}{lll} \hline\hline
           & p-H$_2$CO & o-H$_2$CO \\ \hline
CO         &  0.63    & 0.70    \\
HCO$^+$    &  0.46    & 0.76    \\
CS         &  0.42    & 0.60    \\
SO         &  0.13    & 0.05    \\
HCN        &  0.72    & 0.63    \\
HNC        &  0.57    & 0.74    \\
CN         &  0.57    & 0.81    \\
HC$_3$N    &  0.44    & 0.70    \\[0.5ex]
p-H$_2$CO  & $\ldots$ & 0.92    \\
o-H$_2$CO  &  0.92    & $\ldots$\\ \hline
\end{tabular}\end{center}
\end{table}

\begin{figure}
\resizebox{\hsize}{!}{\includegraphics{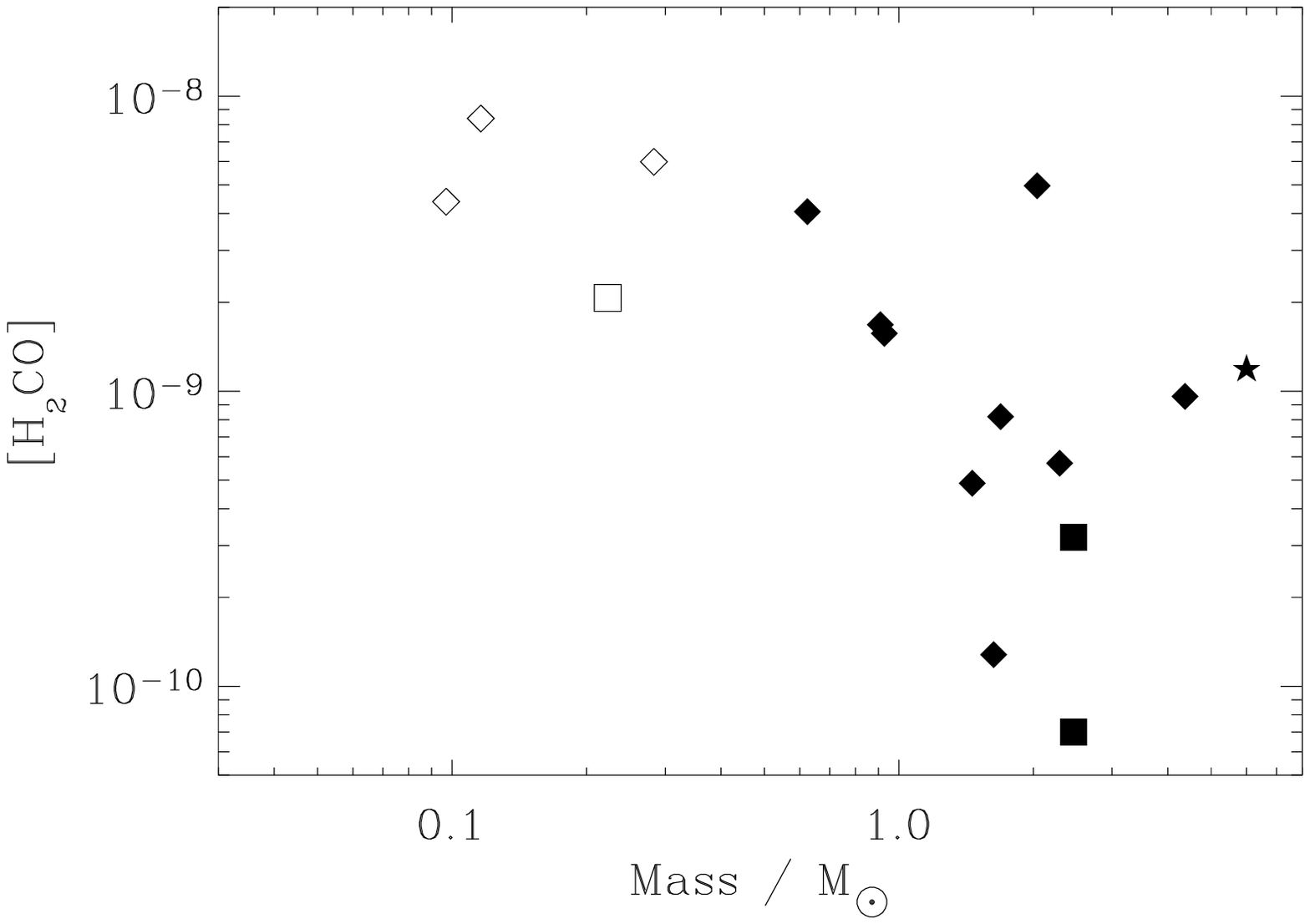}}
\resizebox{\hsize}{!}{\includegraphics{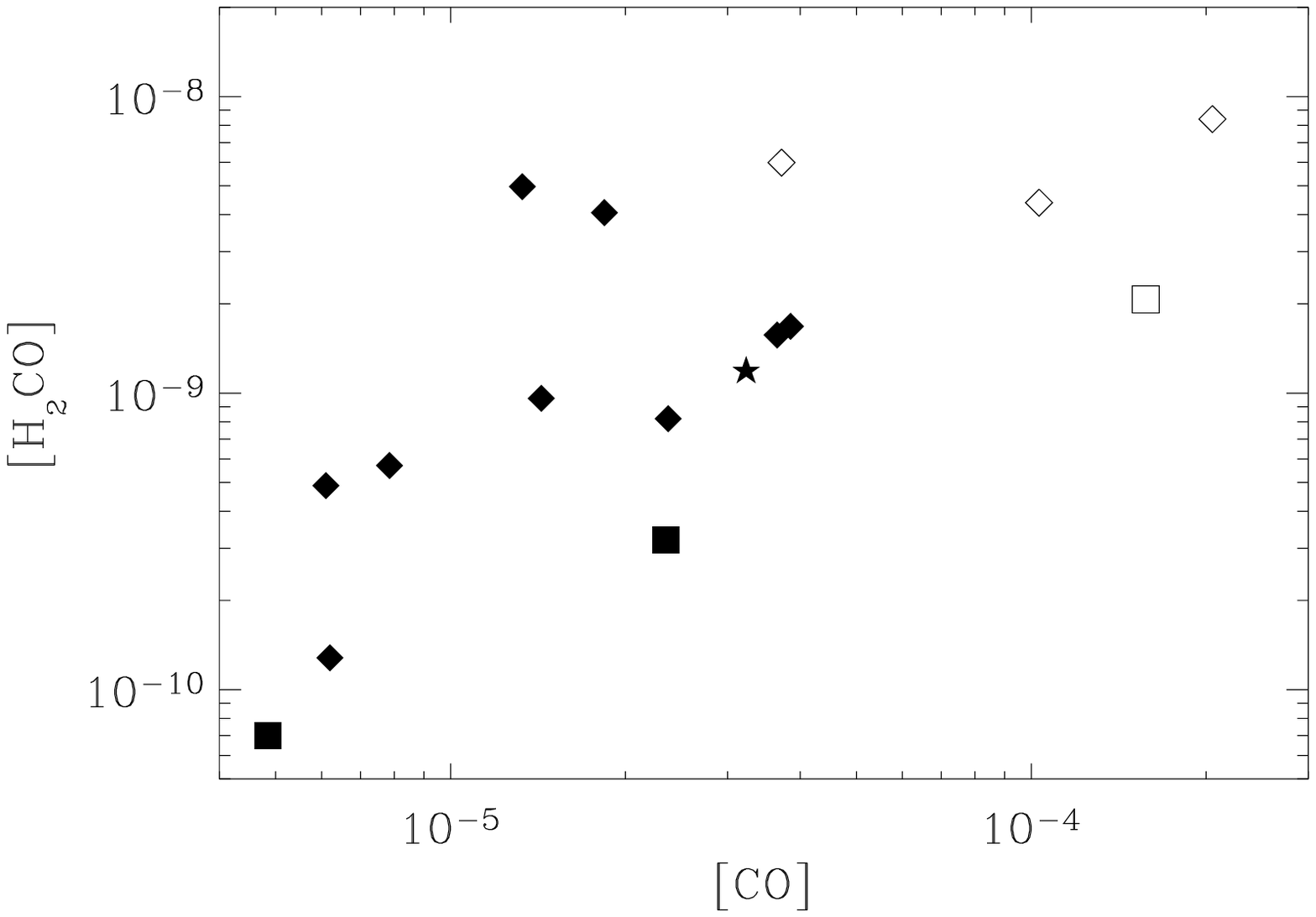}}
\caption{Total H$_2$CO abundance vs. envelope mass (upper panel) and
  CO abundance (lower panel). The envelope mass refers to the mass
  within the radius where the temperature drops to 10~K as estimated
  from the dust radiative transfer models \citep[see][]{jorgensen02}.
  For objects where only the abundance of one of the two H$_2$CO
  species has been constrained, an ortho-para ratio of 1.6 is
  assumed. As in Fig.~\ref{orthopara_abund} the class 0 objects are
  indicated by ``$\blacklozenge$'', the class I objects by
  ``$\lozenge$'' and the pre-stellar cores by ``$\blacksquare$''. The
  class 0 objects \object{VLA1623} and \object{IRAS~16293-2422} have
  been singled out by ``$\square$'' and ``$\bigstar$'',
  respectively.}\label{h2co_mass}
\end{figure}

An important conclusion is the apparent anti-correlation between the
H$_2$CO abundances and envelope mass within the radius where the
temperature drops to 10~K \citep{jorgensen02} as shown in
Fig.~\ref{h2co_mass}.  This is also reflected in the correlations in
Table~\ref{pearson} between H$_2$CO and molecules such as CO whose
abundances decline with increasing mass
\citep{jorgensen02}. \cite{paperii,coevollet} suggested that this
correlation is well explained with a drop-abundance profile, with the
different H$_2$CO abundances reflecting the size of the region over
which freeze-out occurs.

\begin{figure}
\resizebox{\hsize}{!}{\includegraphics{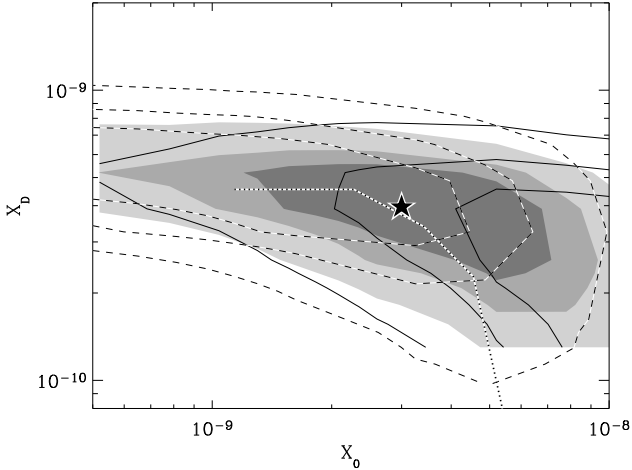}}
\caption{Constraints on H$_2$CO abundances in a ``drop'' model for
NGC~1333-IRAS4A, i.e., a model where the molecule is depleted to
$X_{\rm D}$ in the regions where the density is higher than $n_{\rm
de}$ and temperature lower than $T_{\rm ev}$ constrained from CO
observations. The solid and dashed contours indicate the 1$\sigma$,
2$\sigma$ and 4$\sigma$ confidence levels for p-H$_2$CO and o-H$_2$CO,
respectively whereas the grey-scale contours indicate the confidence
region for the two species combined with an ortho-para ratio of
1.6. The black/white dotted line indicates the best fit relation from
the o-H$_2^{13}$CO lines. The best fit model ($X_0$=3$\times 10^{-9}$, $X_{\rm
D}$=4$\times 10^{-10}$) combining all lines has $\chi^2_{\rm red}$ of 1.1 and
is indicated with the ``$\bigstar$''.}\label{n1333i4a_drop}
\end{figure}

As an illustration, the drop abundance structure is tested for the
H$_2$CO lines toward NGC~1333-IRAS4A. For the H$_2$CO lines we first
constrain the depletion density, $n_{\rm de}$, and desorption
temperature, $T_{\rm ev}$, from observations of the CO lines presented
by \cite{jorgensen02}. The CO data toward NGC~1333-IRAS4A are
well-fitted with depletion by a factor of 50 in the region of the
envelope where the density is higher than 6$\times 10^{5}$~cm$^{-3}$
and temperature lower than 40~K. These constraints are used as input
for the H$_2$CO chemical structure, so that only the overall and
depleted abundances, ($X_0$ and $X_{\rm D}$, respectively) are left as
free parameters. The results of these fits are shown in
Fig.~\ref{n1333i4a_drop}: both ortho and para lines are consistent
with an abundance drop of approximately an order of magnitude. The
$\chi^2$ confidence regions for the two H$_2$CO species agree at the
1$\sigma$ level assuming an ortho-para ratio of 1.6. Also the
o-H$_2^{13}$CO observations agree with those of the main isotopic
lines assuming a $^{12}$C:$^{13}$C ratio of 70. The best fit drop
model has an undepleted abundance $X_0=3\times 10^{-9}$ and an
abundance in the depletion region of $X_{\rm D}=4\times 10^{-10}$. The
reduced $\chi^2$ for this model is 1.1 for 10 fitted lines (including
all ortho, para and H$_2^{13}$CO lines). This suggests that the
variations in H$_2$CO abundances reflect, to first order, the
variations due to freeze-out, with the chemical network subsequently
regulating the abundances. High-resolution interferometer observations
confirm this structure for IRAS~16293-2422 and L1448-C
\citep{hotcorepaper}.

Table~\ref{drop_summary} compares the drop abundance profiles for
NGC~1333-IRAS4A with those of L1448-C and IRAS~16293-2422
\citep{hotcorepaper}. Similar values for $X_0$ are found within a
factor 3 and with the H$_2$CO abundance decreased by an order of
magnitude in the freeze-out zone. Also the maximum constant abundance
for the entire sample is in agreement with this value for $X_0$,
supporting the suggestion that the spread in constant H$_2$CO
abundances reflects the size of the freeze-out zone.
\begin{table}
\caption{Summary of models with varying H$_2$CO abundance
structure.}\label{drop_summary}
\begin{center}\begin{tabular}{lll} \hline\hline
                      & $X_{\rm D}$  & $X_{\rm 0}$ \\ \hline
NGC~1333-IRAS4A       & 4$\times 10^{-10}$   & 3$\times 10^{-9}$   \\
L1448-C$^{a}$         & 1$\times 10^{-9}$    & 1$\times 10^{-8}$   \\
IRAS~16293-2422$^{a}$ & 3$\times 10^{-10}$   & 1$\times 10^{-8}$   \\
Sample$^{b}$          & 7$\times 10^{-11}$   & 8$\times 10^{-9}$   \\[0.5ex]
NGC~1333-IRAS4B$^{c}$ & $<$1$\times 10^{-9}$ & 1$\times 10^{-8}$   \\ \hline
\end{tabular}\end{center}

$^{a}$From \cite{hotcorepaper} fitting both single-dish and
interferometer data. $^{b}$Minimum and maximum abundances for the
entire sample in this paper. $^{c}$Model with abundance jump at
20~K. $X_{\rm D}$ for this source refer to the abundance where
$T<20$~K and $X_0$ to the abundance where $T>20$~K.
\end{table}

\begin{figure}
\resizebox{\hsize}{!}{\includegraphics{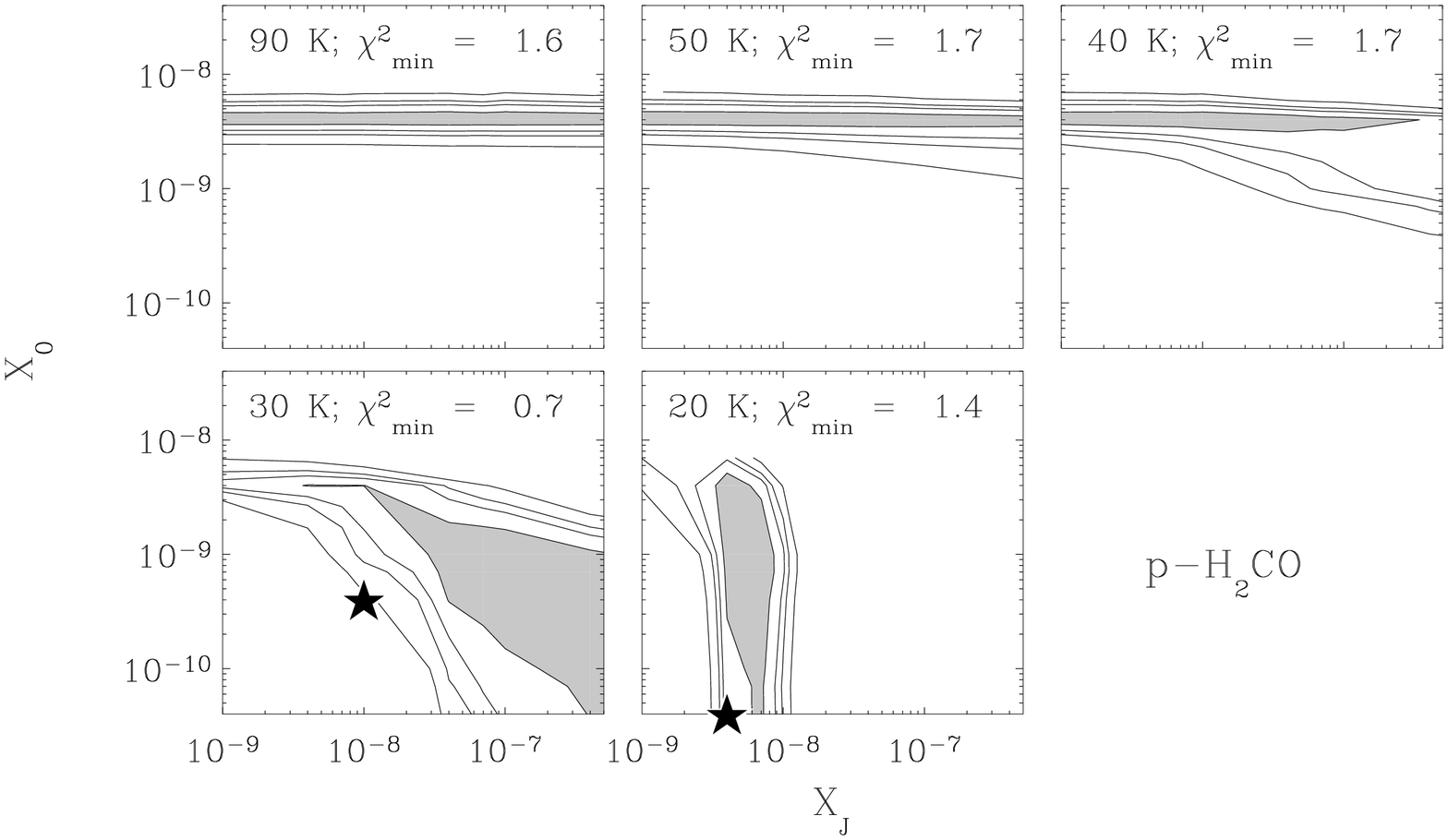}}
\resizebox{\hsize}{!}{\includegraphics{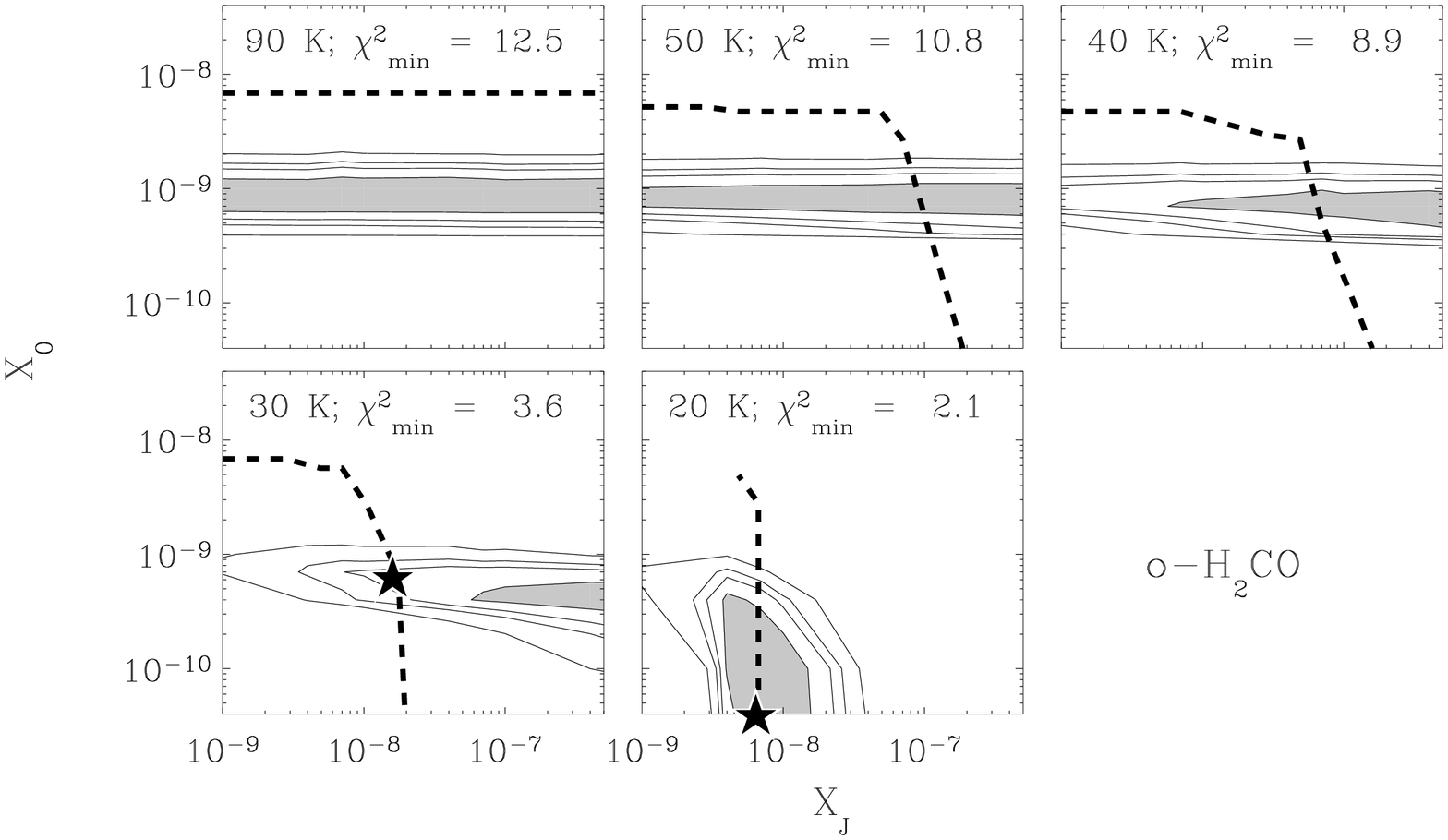}}
\caption{$\chi^2$-confidence contour plots for p-H$_2$CO and o-H$_2$CO
abundances (upper and lower panels, respectively) toward
NGC~1333-IRAS4B. The minimum $\chi^2_{\rm red}$ for the given jump
temperature is indicated in the top of each panel. The dashed lines
indicate the constraints on the abundances from the o-H$_2^{13}$CO
lines. In the $T_{\rm ev}=20$~K and $T_{\rm ev}=30$~K panels the best
fit models combining the constraints for all lines (with minimum
$\chi^2_{\rm red}=1.7$ and 3.8, respectively) have been indicated by
``$\bigstar$''. For the other values of $T_{\rm ev}$, the minimum
$\chi^2_{\rm red} > 5$.}\label{iras4b_h2co_chi2}
\end{figure}

A varying abundance structure is also preferred for NGC~1333-IRAS4B:
as seen from Table~\ref{abund_h2co} and Fig.~\ref{orthopara_abund},
the ortho-para ratio for this particular source is less than 1. This
problem is not alleviated by the introduction of a ``drop'' profile,
which still gives an ortho-para ratio below unity and a poor fit to,
in particular, the ortho-H$_2$CO lines. An abundance increase at low
temperatures, however, does a better job: Fig.~\ref{iras4b_h2co_chi2}
shows models for NGC~1333-IRAS4B with abundance jumps at differing
temperatures. An abundance jump at $T_{\rm ev} \lesssim 30$~K from
$\sim$$10^{-10}$ to $\sim$$10^{-8}$ makes it possible to fit the lines
with an ortho-para ratio above unity, and to bring the abundance
inferred from the o-H$_2^{13}$CO lines in agreement with that of the
o-H$_2^{12}$CO lines. A jump at low temperatures also significantly
improves the best fit for the two species separately. This is
interesting compared with the results of \cite{maret04} who inferred
an abundance enhancement close to 4 orders of magnitude in
NGC~1333-IRAS4B, but with a rather low quality of the fit
($\chi^2_{\rm red}\approx 7$). This suggests that the model with a
jump at temperatures of 90-100~K is not adequate to describe the
abundance structure for NGC~1333-IRAS4B but that other mechanisms such
as the action of the protostellar outflow regulate the H$_2$CO
abundance for this source.
\label{n1333i4b_h2co_txt}

To summarize these discussions: by fitting the ortho and para H$_2$CO
lines independently for the larger sample of sources, the ortho-para
ratio can be constrained statistically to be $1.6\pm 0.3$. The
observed correlations with other species from the survey of
\cite{paperii} suggest that the H$_2$CO abundances are related to the
overall chemical network -- primarily reflecting freeze-out of CO at
low temperatures and high densities. This is further illustrated by
fits to the NGC~1333-IRAS4A H$_2$CO data, which are improved with a
``drop abundance'' profile. In these models, the H$_2$CO abundance is
decreased by an order of magnitude in the cold freeze-out zone. No
jump in abundance in the innermost ($T\gtrsim 90$~K) region is needed,
although NGC~1333-IRAS4B is best fit with an abundance increase where
$T \gtrsim 20-30$~K. Observations of higher excitation lines are
needed to constrain jumps in the hot core region. The NGC~1333-IRAS4B
data suggest that for specific sources, other effects, such as the
impact of the outflow, may play a role in defining the H$_2$CO
abundance structures.

\subsection{CH$_3$OH and CH$_3$CN}
The upper limits for the constant CH$_3$OH abundances found for the
sources in sample range from a few$\times 10^{-10}$ to
$\sim$10$^{-9}$. The upper limits are typically a factor of a few
below the abundances determined from lower excitation CH$_3$OH lines
by \cite{buckle02}. However, since their abundances were calculated
relative to CO, for which abundances are lower by an order of
magnitude than the canonical value in the class 0 objects, the upper
limits from this paper and the results of \cite{buckle02} are still
consistent. \citeauthor{buckle02} also found a slight decrease in
CH$_3$OH abundance with bolometric temperature (i.e., lower CH$_3$OH
abundance in the class I objects). This may again be a result of the
abundance calculated relative to CO, since the CO abundance varies
with envelope mass \citep{jorgensen02}. A similar trend was seen for
the sulfur-bearing species \citep[][see discussion in J{\o}rgensen et
al. 2004d]{buckle03}.

As noted by \cite{maret04ch3oh}, the CH$_3$OH data for NGC~1333-IRAS2
and -IRAS4B (see also Table~\ref{abund_ch3oh}) cannot be modeled with
constant abundances. In the best cases such models give $\chi^2_{\rm
red}\approx 10-20$. NGC~1333-IRAS4A also shows mediocre fits for each
of the A and E-type species with constant abundances, but still better
than the two other sources ($\chi^2_{\rm red}\approx 3-4$).

Our analysis uses the new collisional rate coefficients for CH$_3$OH
recently published by \cite{pottage04}. Compared to the old rate
coefficients, the derived line intensities vary in certain cases by up
to 50\%. However, no systematic trends are seen and therefore the
derived abundance structures are unchanged. Still, this example
illustrates that the derived abundances -- especially when based on
constraints from only a few lines -- may be uncertain by up to a
factor of 2 due to uncertainties in the collisional data alone
\citep[see also][]{schoeier03radex}.\label{ch3oh_newrates}

\begin{figure}
\resizebox{\hsize}{!}{\includegraphics{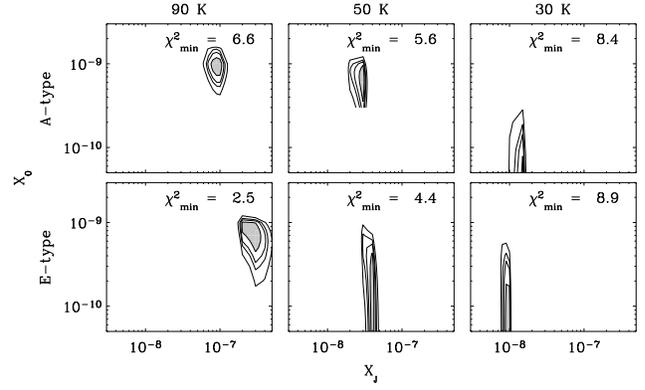}}
\caption{$\chi^2$-confidence contours for jump models with varying
$T_{\rm ev}$ for CH$_3$OH lines toward NGC~1333-IRAS2. The minimum
$\chi^2_{\rm red}$ is given in the upper right corner of each
panel.}\label{iras2_ch3oh_chi2}
\end{figure}

\begin{figure}
\resizebox{\hsize}{!}{\includegraphics{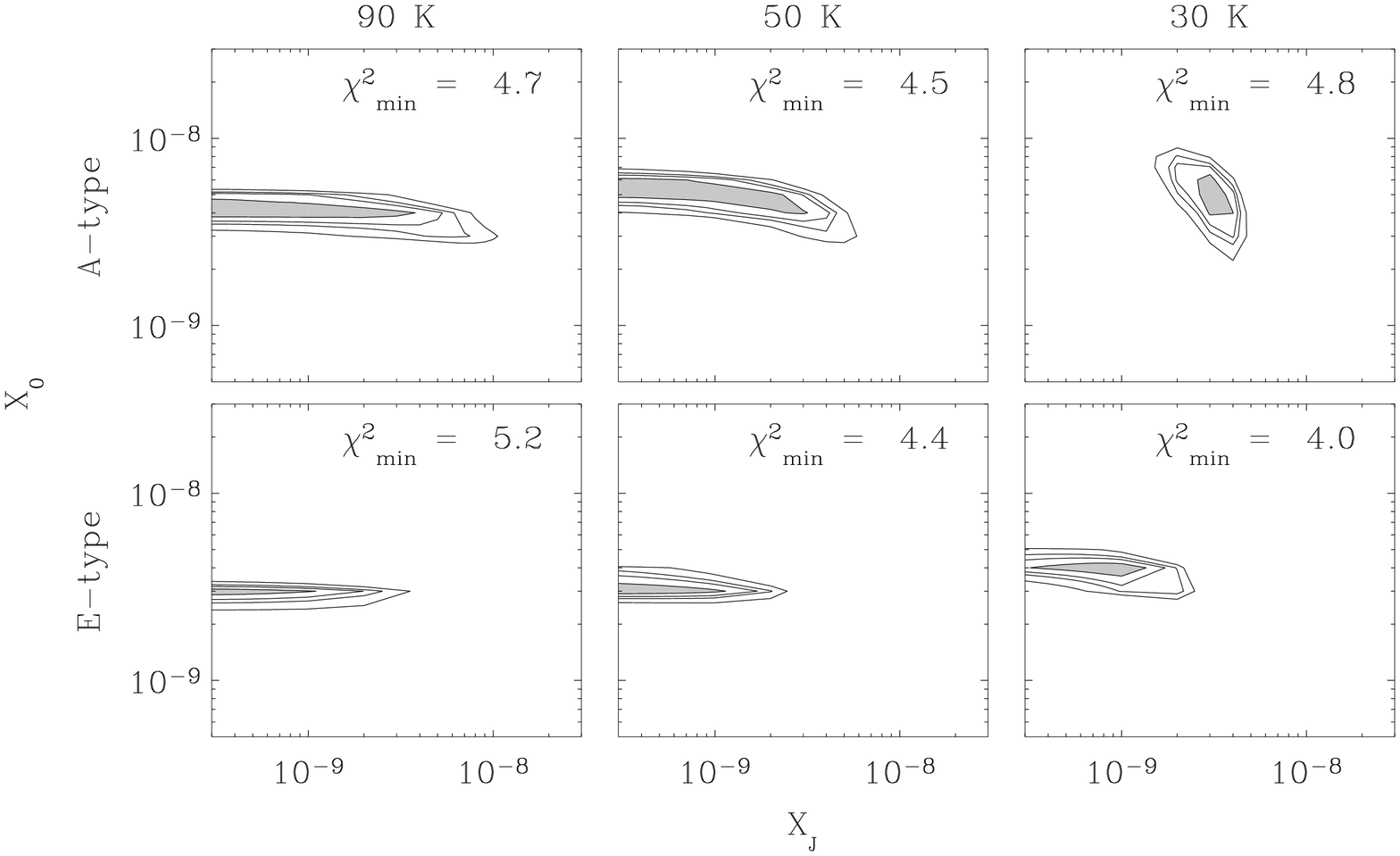}}
\caption{As in Fig.~\ref{iras2_ch3oh_chi2} for CH$_3$OH lines toward
NGC~1333-IRAS4A.}\label{iras4a_ch3oh_chi2}
\end{figure}

\begin{figure}
\resizebox{\hsize}{!}{\includegraphics{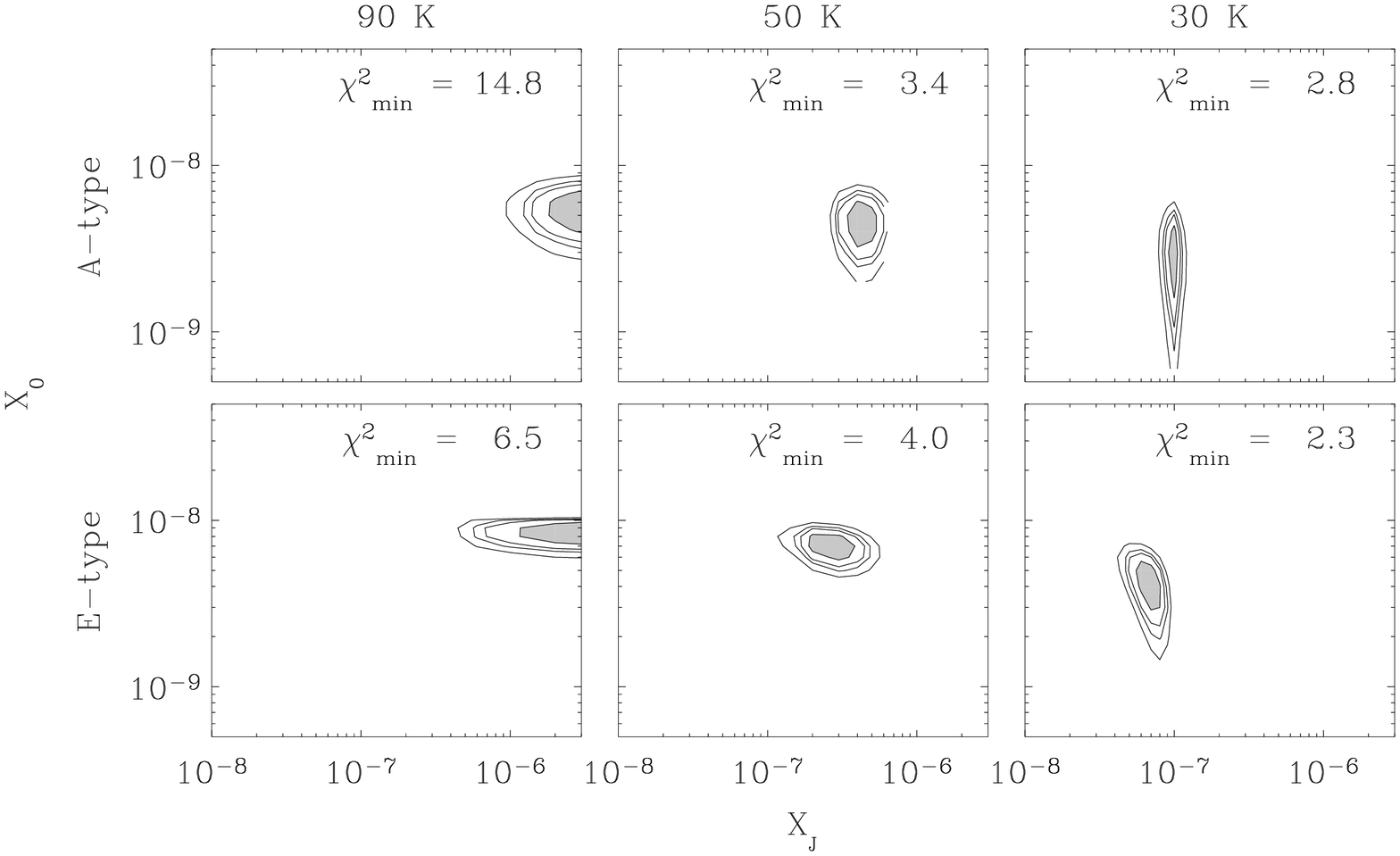}}
\caption{As in Fig.~\ref{iras2_ch3oh_chi2} for CH$_3$OH lines toward
NGC~1333-IRAS4B.}\label{iras4b_ch3oh_chi2}
\end{figure}

The poor fits can be improved by including evaporation of grain ice
mantles at temperatures $\gtrsim 90$~K
\citep{ceccarelli00a,schoeier02,maret04,maret04ch3oh}. To test this, a
step function for the abundance was introduced, with a jump in
abundance from $X_0$ in the exterior to $X_{\rm J}$ in the interior at
a radius corresponding to a specific temperature $T_{\rm ev}$. Models
were run for $T_{\rm ev}=$~30, 50 and 90~K for each of the three
NGC~1333 sources. Table~\ref{jump_chi2} gives the best fit models and
Fig.~\ref{iras2_ch3oh_chi2}-\ref{iras4b_ch3oh_chi2} show the derived
$\chi^2$ confidence plots for each of the temperatures and for each of
the sources. They clearly show different behavior: NGC~1333-IRAS2 is
nicely fit with a jump at 90~K, whereas NGC~1333-IRAS4B is much better
fit with a jump at 30~K. For NGC~1333-IRAS4A the models suggest a best
fit for a constant or ``anti-jump'' abundance structure.

It is also noteworthy that the best fits of both A- and E-type species
are obtained for similar abundances at the best fitting evaporation
temperatures. As for the H$_2$CO ortho-para species, no transitions
between the A- and E-type CH$_3$OH levels are expected. The E/A-type
abundance ratio, however, only varies from 0.69, corresponding to
thermalization at 10~K, to unity in the high temperature limit
\citep{friberg88}. The abundances are therefore expected to be similar
for the best fit models, which appears to be the case as illustrated
in Fig.~\ref{iras2_ch3oh_chi2}-\ref{iras4b_ch3oh_chi2}.
\begin{table*}
\caption{Derived CH$_3$OH and CH$_3$CN abundances assuming abundance
jumps in the inner $T > T_{\rm ev}$ regions.}\label{jump_chi2}
\begin{center}\begin{tabular}{llllll} \hline\hline
   & $T_{\rm ev}$ & Species & $X_{\rm J}$ & $X_{\rm 0}$ & $\chi^2_{\rm red}$ \\ \hline
\multicolumn{6}{c}{CH$_3$OH} \\ 
IRAS2  & 90~K & A-type (8)  & 9$\times 10^{-8}$    & 1$\times 10^{-9}$      & 6.6  \\
       &      & E-type (13) & 3$\times 10^{-7}$    & 7$\times 10^{-10}$     & 2.5  \\[2.0ex]
IRAS4A & 50~K & A-type (7)  & $\le$5$\times 10^{-9}$ & 4$\times 10^{-9}$    & 4.5$^a$ \\
       &      & E-type (7)  & $\le$2$\times 10^{-9}$ & 3$\times 10^{-9}$    & 4.4$^a$ \\[2.0ex]
IRAS4B & 30~K & A-type (7)  & 1$\times 10^{-7}$    & 3$\times 10^{-9}$      & 2.8  \\
       &      & E-type (7)  & 9$\times 10^{-8}$    & 3$\times 10^{-9}$      & 2.7  \\[2.0ex]
IRAS~16293-2422$^{b}$ & 90~K & A+E-type (23) & 1$\times 10^{-7}$ & 6$\times 10^{-9}$ & 1.2  \\[2.0ex]
\multicolumn{6}{c}{CH$_3$CN} \\
IRAS2  & 90~K & A-type      & 7$\times 10^{-9}$    & $<3$$\times 10^{-11}$  & 1.4 \\ \hline
\end{tabular}\end{center}

$^a$No strong constraints exist on the evaporation temperature for
NGC~1333-IRAS4A in accordance with the conclusion that this source is
well-fitted by a constant abundance throughout the envelope (see
Fig.~\ref{iras4a_ch3oh_chi2}). $^b$Results for IRAS~16293-2422 from
\cite{schoeier02} assuming the abundances of the A- and E-type
CH$_3$OH to be identical.
\end{table*}

\begin{figure}
\resizebox{\hsize}{!}{\includegraphics{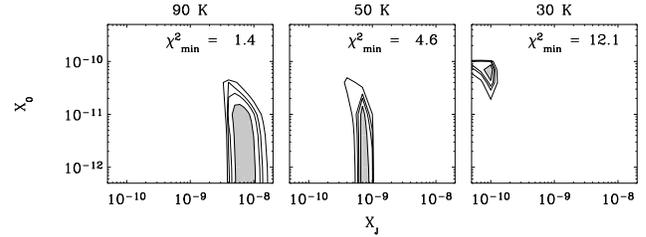}}
\caption{$\chi^2$-confidence contours for jump models with varying
$T_{\rm ev}$ for CH$_3$CN lines toward NGC~1333-IRAS2. The minimum
$\chi^2_{\rm red}$ is given in the upper right corner of each
panel.}\label{iras2_ch3cn_chi2}\label{iras2_ch3cn}
\end{figure}

For NGC~1333-IRAS2 similar jump models were run for CH$_3$CN, and the
best fit abundance is shown in Fig.~\ref{iras2_ch3cn}. Interestingly
the CH$_3$CN lines also give an abundance jump of approximately two
orders of magnitude at 90~K, similar to what is found for the CH$_3$OH
lines. Again jumps at lower temperatures are not favored for this
source. The value of $X_{\rm J}$ for NGC~1333-IRAS2 is comparable to
that derived for IRAS~16293-2422 by \cite{schoeier02} and
\cite{cazaux03}. There is no strong evidence to suggest that the
chemistry of this molecule is significantly different in the innermost
envelopes around the objects discussed in this paper compared to
IRAS~16293-2422.

\section{Discussion}\label{discussion}
\subsection{Hot core vs. outflow}
As discussed in Sect.~\ref{h2co_model}, the H$_2$CO abundances are
found to be consistent with constant abundances throughout the
envelopes for most sources. The CH$_3$OH results for the three
NGC~1333 sources, in contrast, imply abundance variations, but these
occur over significantly different regions of the envelope. The three
objects have rather similar density and temperature profiles and the
observed differences therefore suggest other causes for the abundance
enhancements than passive heating of the envelope material.

\begin{figure}
\resizebox{\hsize}{!}{\includegraphics{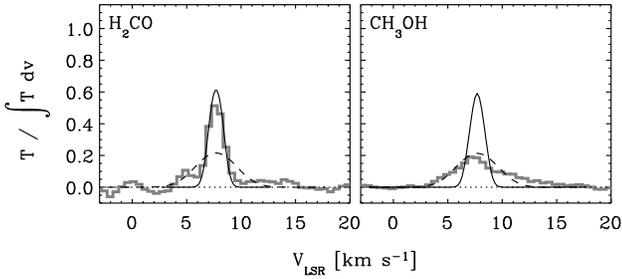}}
\caption{Observed and modeled spectra of H$_2$CO $5_{05}-4_{04}$ and
  CH$_3$OH $7_{-1}6_{-1}$-E lines toward NGC~1333-IRAS2. The two
  models shown use turbulent line broadening of 0.8~km~s$^{-1}$, which
  can also account for, e.g., the C$^{18}$O lines modeled in
  \cite{jorgensen02} (solid line), and 2.5~km~s$^{-1}$ (dashed line),
  respectively. Note how the two lines probe significantly different
  velocity fields in the envelope.}\label{h2co_ch3oh_velocity}
\end{figure}

An important clue comes from the differing line profiles of H$_2$CO
and CH$_3$OH. Fig.~\ref{h2co_ch3oh_velocity} compares the profiles for
the H$_2$CO $5_{05}-4_{04}$ and CH$_3$OH $7_{-1}-6_{-1}$-E lines
toward NGC~1333-IRAS2. The CH$_3$OH line is significantly broader,
with a width of 4~km~s$^{-1}$ compared to the 1.5~km~s$^{-1}$ for the H$_2$CO
line. Part of this could be due to differences in the thermal
broadening if CH$_3$OH probes warmer gas much deeper in the
envelope. The radiative transfer models, however, take this explicitly
into account and it is concluded that a significantly higher turbulent
broadening is required to model the observed CH$_3$OH lines compared
with those of H$_2$CO and other species. This suggests that the
CH$_3$OH lines probe a different part of the envelope than H$_2$CO and
the species discussed by \cite{paperii}.

One explanation for the enhancements and profiles of CH$_3$OH could be
the impact of the outflows on the inner envelopes. This was suggested
to be the case for NGC~1333-IRAS4A and IRAS4B by \cite{blake95}, who
noted that the abundance enhancements can occur in the outflow shear
zones probed by the very broad, kinematically distinct CS and CH$_3$OH
lines. Large enhancements of CH$_3$OH are seen in the shocks driven by
the outflows from a number of protostellar sources where the shock is
well separated from the central protostar including L1448-C, L1157 and
NGC~1333-IRAS2
\citep[e.g.,][]{bachiller95,bachiller97,i2art}. CH$_3$OH is one of the
molecules that shows the largest jumps in abundances between cold and
warm gas and therefore traces more clearly the origin of the abundance
enhancements.

\begin{figure}
  \resizebox{\hsize}{!}{\includegraphics{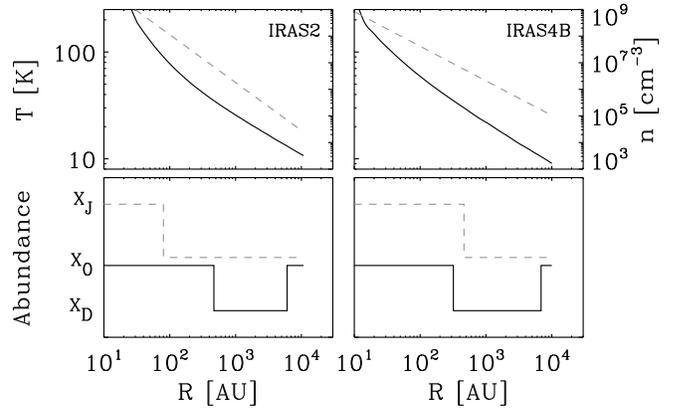}}
\caption{Comparison of the inferred temperature, density and abundance
  structures in NGC~1333-IRAS2 (left) and NGC~1333-IRAS4B (right). The
  upper panels show the temperature (black solid line) and density
  (grey dashed line) profiles for each source. The lower panels show
  the inferred CO drop abundance (black solid line) and CH$_3$OH
    jump abundance (grey dashed line) profiles \citep[][this
      paper]{coevollet}.}\label{abundstrucfig} \end{figure}

An outflow scenario could explain the relatively low desorption
temperature found for the CH$_3$OH lines as well as the fits to the
H$_2$CO lines for NGC~1333-IRAS4B discussed in
Sect.~\ref{n1333i4b_h2co_txt}. If these species are desorbed from
grains due to the action of the outflow, they could thermalize at
temperatures closer to that of the envelope, i.e., lower than in the
small ``hot core'' region. NGC~1333-IRAS4A is remarkable since the
fits to the CH$_3$OH lines do not require a jump in abundance. This
difference from NGC~1333-IRAS4B could be caused by the differences in
the outflow morphologies: \cite{difrancesco01} imaged the
NGC~1333-IRAS4 region at high resolution and found that the outflows
are probed by the wings of H$_2$CO and CS.  The images show that the
IRAS4B outflow is more compact with emission over a
$\approx$~15\arcsec\ region comparable to the 350~GHz JCMT single-dish
beam. The morphology thus suggests that enhancements for the
NGC~1333-IRAS4B outflow occur on scales where the single-dish
observations are the most sensitive. Also the SiO $J=$5--4 line at
217.1~GHz is only detected toward NGC~1333-IRAS4A and not IRAS4B. This
indicates that the CH$_3$OH enhancements in IRAS4B are not directly
associated with a high velocity shock but more likely result from the
shear between the envelope and outflow. Note also that direct images
of outflow induced shocks
\citep{bachiller97,garay00,bachiller01,i2art} indicate that CH$_3$OH
does not survive at the highest shock speeds, which could further
explain the differences between IRAS4A and IRAS4B. In contrast, the
impact of the outflow in NGC~1333-IRAS2 is on much smaller scales,
more heavily diluted in the single-dish beam.
Fig.~\ref{abundstrucfig} compares the inferred abundance structures
for NGC~1333-IRAS2 and -IRAS4B.  

\subsection{CS $J=10-9$ and HDO as dense gas probes}\label{cs_hdo}
\begin{figure}
\resizebox{\hsize}{!}{\includegraphics{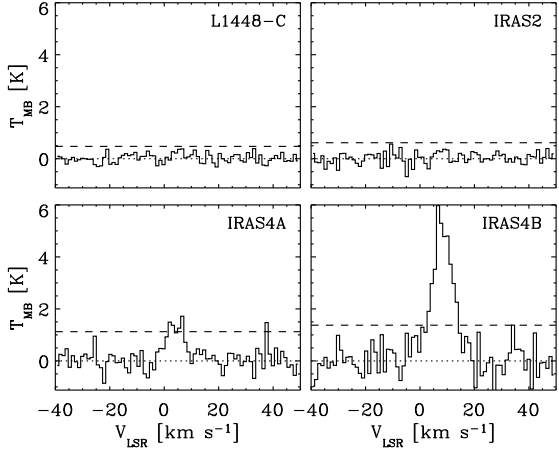}}
\caption{CS $J=10-9$ observations of L1448-C and the
three NGC~1333 sources. In each plot the dashed line indicates the
3$\sigma$ detection limit. Lower 6 panels: Comparison between CS lines
probing different excitation conditions in the envelopes, i.e., depths
of NGC~1333-IRAS2 and -IRAS4B. The lowest excitation 3--2 lines are
from IRAM 30~m observations the remainder from JCMT
observations.}\label{cs109_fig}
\end{figure}

\begin{figure}
\resizebox{\hsize}{!}{\includegraphics{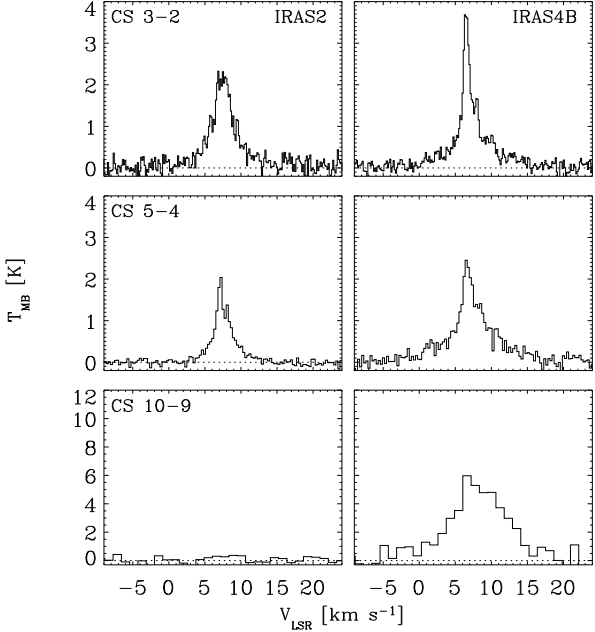}}
\caption{Comparison between CS lines probing different excitation
conditions in the envelopes, i.e., depths of NGC~1333-IRAS2 and
-IRAS4B. The lowest excitation 3--2 lines are from IRAM 30~m
observations the remainder from JCMT observations.}\label{csline_comparison}
\end{figure}

Further support for the importance of the outflows comes from
observations of high excitation CS $J=10-9$ and HDO lines obtained
with the JCMT. For CS $J=10-9$, broad lines (FWHM $\approx 8$~km~s$^{-1}$)
are detected toward both NGC~1333-IRAS4A and IRAS4B
(Fig.~\ref{cs109_fig}), which lack the central narrow peak seen for
the lower excitation lines (Fig.~\ref{csline_comparison}). Although,
the absolute calibration may be somewhat uncertain, the CS $J=10-9$
line is approximately 5 times stronger toward IRAS4B than IRAS4A,
consistent with the above conclusion that the dense outflow gas fills
a larger fraction of the beam for IRAS4B. This is in contrast to the
lower excitation CS lines reported, e.g., by \cite{blake95} and
\cite{paperii}, which support a more compact origin of the CS outflow
emission in IRAS4B than IRAS4A.

Table~\ref{cs_comparison} compares the predictions for the CS $J=10-9$
line intensity assuming CS abundances in the quiescent envelope from
the models of \cite{paperii}. The envelope models predict significantly
less emission than observed for IRAS4A and IRAS4B, but the
non-detection toward IRAS2 is consistent within the 3$\sigma$ noise
level. Note that the CS intensities in \cite{paperii} were found by
integration over $\pm 2$~km~s$^{-1}$ from the systemic velocity: for the CS
$J=10-9$ lines the emission integrated over this velocity range only
contributes 20--25\% of the total integrated emission and is still
underestimated by the envelope models, especially for IRAS4B. Again
this suggests that the observed CS $J=10-9$ emission probes different
material than traced in the bulk envelope material. 

Another simple estimate can be made assuming that the CS emission
comes from a medium with a constant density and kinetic
temperature. Using a non-LTE escape probability code, \emph{Radex}
\citep{jansen94,schoeier03radex}, the CS column density is estimated
adopting a density of $3\times 10^6$~cm$^{-3}$ and kinetic temperature
of 100~K \citep{blake95}, consistent with the intensity of the wing
emission from the CS 5--4 and 7--6 lines from \cite{paperii}. A high
column density of $\sim 5\times 10^{14}$~cm$^{-2}$ is needed to
produce the observed CS 10--9 emission. Even at such high column
densities, the emission is found to be optically thin. This column
density is an order of magnitude larger than found from the lower
excitation lines by \cite{blake95} and corresponds to $\approx 5-10$\%
of the estimated CO abundance in the outflowing material based on the
CO column density calculated by \citeauthor{blake95}.
\begin{table}
\caption{Observed and predicted line intensities $(\int T_{\rm
MB}\,{\rm d}v)$ for CS $J=10-9$ for the three sources in
NGC~1333.}\label{cs_comparison}
\begin{center}\begin{tabular}{lll} \hline\hline
Source  & $I_{\rm mod}$ [K km~s$^{-1}$]$^{a}$  & $I_{\rm obs}$ [K km~s$^{-1}$]$^{b}$  \\ \hline
IRAS2   & 1.8 & $<2$ \\
IRAS4A  & 1.7 & 12 (3)  \\
IRAS4B  & 1.1 & 51 (11) \\ \hline
\end{tabular}\end{center}

$^{a}$Predicted CS $J=10-9$ line intensity adopting best fit abundances
for each source from \cite{paperii}. $^{b}$Total line emission from
Gaussian fits (IRAS4A and IRAS4B) or as 3$\sigma$ upper limit
(IRAS2). For IRAS4A and IRAS4B the number in parenthesis indicate the
line emission integrated over $\pm 2$~km~s$^{-1}$ from the systemic
velocity.
\end{table}

\begin{figure}
  \resizebox{\hsize}{!}{\includegraphics{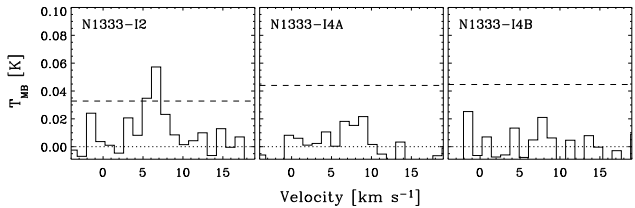}}
\caption{Observations of the HDO $2_{11}-2_{12}$ line at 241.5616~GHz
toward the three NGC~1333 sources observed in the CH$_3$OH 5--4
setting. HDO is only detected toward NGC~1333-IRAS2. In each plot the
dashed line indicates the 3$\sigma$ detection limit.}\label{hdofig}
\end{figure}

In contrast to the CS $J=10-9$ lines, HDO $2_{11}-2_{12}$ is detected
only toward NGC~1333-IRAS2 and not IRAS4A and IRAS4B (see
Fig.~\ref{hdofig}). Like CS 10--9 this line probes the warm, dense gas
with an upper level energy of 90~K. The observed line is narrow (FWHM
of $\approx 2.5$~km~s$^{-1}$) compared to the $\approx 8$~km~s$^{-1}$ for the CS
$J=10-9$ and $\approx 4$~km~s$^{-1}$ for the CH$_3$OH lines toward IRAS4A
and IRAS4B. This suggests that this line has its origin in a ``hot
inner region'' of the NGC~1333-IRAS2 envelope, although relation to
the small-scale outflow \citep[as seen in high resolution maps
  by][]{n1333i2art} cannot be ruled out. Enhancements of HDO by up to
a factor of 10 were derived for the IRAS~16293-2422 outflow by
\cite{stark04}, but \cite{parise05} show that observations of a larger
number of HDO transitions are better explained by an abundance
enhancement by two orders of magnitude in the innermost quiescent
envelope. Observations of more transitions will be needed before
either scenario can be confirmed or ruled out for
NGC~1333-IRAS2. Since both CH$_3$OH and CH$_3$CN observations indicate
the presence of warm gas with abundance jumps in the inner ($T< 90$~K)
region, NGC~1333-IRAS2 still seems the best candidate for further
comparative studies of the passively heated, hot inner regions of
low-mass protostellar envelopes.  

\subsection{Comparison with IRAS~16293-2422}
The above results again raise the question what causes the richness of
lines in IRAS~16293-2422 (even taking into its smaller distance and
higher luminosity compared to, e.g., the NGC~1333 sources) since its
abundances in the outer envelope are comparable to those of other
sources. One explanation may be that its small-scale physical
structure is significantly different from that of the remaining
sources. The central binary system has a separation of 8-10$\arcsec$
and affects the material in the envelope with emission from the
various species centered around one or both components
\citep{mundy92,hotcorepaper}. Furthermore the circumbinary envelope
appears to have an inner cavity of size comparable to the binary
separation. It may be that this relatively wide ($\sim$1000~AU) binary
pushes material to larger distances where it can more easily be
observed through single-dish observations with
$\sim$10--15\arcsec\ beams. Alternatively, the circumbinary envelope
may be heated on larger scales through each of the relatively luminous
components compared with that expected from a single component in a
simple spherical envelope. Finally, interferometer maps also show
velocity gradients indicating that the outflow affects the envelope
material close to the central protostar.  It is possible that the
outflow processing leads to abundance enhancements of the organic
species on larger scales than the passively heated hot core, thus
providing a larger filling factor of the single-dish beam.

\section{Conclusions} We have presented an analysis of H$_2$CO and
CH$_3$OH line observations for a sample of 18 pre- and protostellar
cores which have previously been studied through continuum and line
observations and detailed radiative transfer modeling
\citep{jorgensen02,paperii}. These results complement the results by
\cite{maret04,maret04ch3oh} for a subset of sources. In addition,
observations and limits for high excitation CS $J=10-9$ transitions
and lines of HDO and CH$_3$CN are presented for a subset of
sources. Molecular abundances are derived through Monte Carlo line
radiative transfer and compared to the results from the survey of
\cite{paperii}. The main conclusions are: \begin{itemize} \item The
  H$_2$CO data of most sources can be well-fitted by constant
  abundances throughout their envelopes with an ortho-para ratio of
  1.6$\pm 0.3$. This implies thermalization of H$_2$CO at low
  temperatures, e.g., on grain ice-mantles. Higher angular resolution
  data are needed to constrain the presence of any abundance jumps in
  the inner warm envelopes.  \item The H$_2$CO abundances are related
  to the chemical network of the other species indicating that the
  same processes regulate their abundances. As an example the H$_2$CO
  data for NGC~1333-IRAS4A are well-fit by ``drop abundance'' profiles
  with a decrease in abundance from a~few~$\times 10^{-9}-10^{-8}$ to
  a~few~$\times 10^{-10}$ in a limited part of the envelope, bounded
  inwards by the desorption temperature and outwards by a density
  corresponding to the lifetime of the core. A counter example is
  provided by NGC~1333-IRAS4B, where an abundance increase is only
  needed where the temperature rises above 20--30~K with no
  enhancement in the outermost regions. This indicates that for some
  sources other effects, such as the impact of an outflow, may be
  important for regulating the H$_2$CO abundances.  \item The upper
  limits to the CH$_3$OH abundances for the entire sample are a
  few$\times 10^{-10}$--10$^{-9}$. These results are consistent with
  the abundances determined by \cite{buckle02} from lower excitation
  lines.
\item CH$_3$OH observations for NGC~1333-IRAS2 and NGC~1333-IRAS4B
  require abundance jumps of at least two orders of magnitude at 90~K
  and 30~K, respectively. Together with the significantly broader
  lines of CH$_3$OH compared with H$_2$CO and other species, this
  suggests that the abundance increase for IRAS4B, in particular, is
  due to a compact outflow interacting with the nearby envelope. This
  is further supported by the broad high frequency CS $J=10-9$ line,
  detected very strongly toward this source. HDO and CH$_3$CN, in
  contrast, are only observed for NGC~1333-IRAS2 - possibly reflecting
  the presence of a passively heated, warm inner region for this
  source where molecules can evaporate. The CH$_3$CN data for
  NGC~1333-IRAS2 also require a jump in abundance at 90~K by about two
  orders of magnitude.  \end{itemize}

This paper reinforces the importance of identifying (if possible)
unique chemical tracers of material heated ``passively'' by a central
protostar and by shocked material in outflows. Even relatively high
excitation lines from single-dish observations (such as the CS
$J=10-9$ lines) may be affected by the outflow and can provide a good
indication of the filling factor of dense shocked material. Future
Herschel-HIFI observations may provide additional tests of the
chemical structure by observations of high frequency lines, but the
fact that shock and envelope chemistry are spatially unresolved
may be problematic for these large beam data. Possibly the best way of
distinguishing the different chemical scenarios will be through high
angular resolution, high excitation observations with facilities such
as the SMA and ALMA - or from studies of outflows well-separated from
the central protostar as, for example, done in \cite{bachiller97} and
\cite{i2art}. Better knowledge about the physical properties of the
inner envelopes will also be important, since the adopted envelope
models are extrapolations from the larger-scale observations of the
cold dust in the outer envelope. Infrared observations with the
Spitzer Space Telescope can place better constraints there.

\begin{acknowledgements}
The authors thank Sebastien Maret and Cecilia Ceccarelli for
interesting discussions and communicating their results prior to
publication. The referee is thanked for detailed comments that helped
clarifying the paper. The research of JKJ is funded by a NOVA network
2 Ph.D. stipend. FLS acknowledges support from the Swedish Research
Council. Astrochemistry research in Leiden is supported by an NWO
Spinoza grant.
\end{acknowledgements}

\clearpage
\appendix
\section{Line widths from gaussian fits}\label{widthfits}
Table~\ref{h2co_widths}--\ref{ch3oh_widths} lists the line widths
(FWHM) for the observations where Gaussians could be fitted. For the
lower excitation H$_2$CO observations not listed here the lines were
integrated over $\pm$2~km~s$^{-1}$. 
\begin{table}[!hbt]
\caption{Line widths (FWHM in km~s$^{-1}$) for H$_2$CO lines (see also \cite{maret04}).}\label{h2co_widths}
\begin{tabular}{lll} \hline\hline
Source            & $\Delta v$($5_{05}-4_{04}$) & $\Delta v$($5_{15}-4_{14}$) \\ \hline
       L1448-I2   & 1.5    & 0.96   \\ 
       L1448-C    & 1.1    &  --$^b$ \\
       N1333-I2   & 1.9$^a$ & --$^b$ \\
       N1333-I4A  & 3.5$^a$ & 4.1$^a$ \\
       N1333-I4B  & 3.1$^a$ & 4.0$^a$ \\
          L1527   & 1.0    & 0.9    \\
        VLA1623   & 0.63   & --$^b$ \\ 
           L483   & 0.93   & 1.4    \\ 
           L723   & 1.6    & 2.3    \\
          L1157   & 0.52   & 1.1    \\ 
          CB244   & 1.4    & 1.8    \\ 
          L1551   & 1.6    & 1.3    \\ 
          L1489   &$\ldots$& 2.0$^a$ \\ 
          TMR1    &$\ldots$& 1.9    \\ \hline
\end{tabular}

Notes: $^a$Non-Gaussian line profile. Intensity in
Table~\ref{h2co_int} refers to line integrated over
$\pm$2~km~s$^{-1}$. Line width defined as $\Delta v = \int T\,{\rm d}v
/ (1.064\, T_{\rm peak})$ where the integral is over the total line
profile and $T_{\rm peak}$ is the temperature at the line peak.
$^{b}$Insufficient spectral resolution to estimate width.
\end{table}
\begin{table}[!hbt]
\caption{CH$_3$OH line widths (FWHM in km~s$^{-1}$) for sources in NGC~1333.}\label{ch3oh_widths}
\begin{center}\begin{tabular}{lllll} \hline\hline
Line & Frequency &             IRAS2  & IRAS4A & IRAS4B \\ \hline
\multicolumn{5}{c}{$5_K-4_K$ band; E-type} \\ \hline
    +0E &           241.7002 & 3.5     & 6.0$^a$ & 3.4$^a$ \\
  $-$1E & \phantom{241}.7672 & 3.0$^a$ & 6.2$^a$ & 3.4$^a$ \\
  $-$4E & \phantom{241}.8132 & 4.9     &$\ldots$ &$\ldots$\\
    +4E & \phantom{241}.8296 & 2.1     &$\ldots$ &$\ldots$\\
    +3E & \phantom{241}.8430 & --      & --      & --   $\dagger$  \\
  $-$3E & \phantom{241}.8523 & 3.9     &$\ldots$&$\ldots$\\
    +1E & \phantom{241}.8790 & 3.4     & 6.0     & 3.4    \\
$\pm$2E & \phantom{241}.9044 & 3.4     & 6.9     & 3.5    \\ \hline
\multicolumn{5}{c}{$5_K-4_K$ band; A-type} \\ \hline
    +0A &           241.7914 & 2.6$^a$ & 5.7$^a$ & 3.2$^a$ \\
$\pm$4A & \phantom{241}.8065 & 3.2     &$\ldots$&$\ldots$\\
$\pm$3A & \phantom{241}.8329 & 3.4     & 6.0    & 2.8    \\
  $-$2A & \phantom{241}.8430 & --      & --     & --$\dagger$  \\
    +2A & \phantom{241}.8877 & 3.2     & 6.0    & 4.3    \\ \hline
\multicolumn{5}{c}{$7_K-6_K$ band; E-type} \\ \hline
  $-$1E &           338.3446 & 4.5$^a$ & 7.7$^a$ & 4.4$^a$ \\
  $-$4E & \phantom{338}.5040 & 3.3     &$\ldots$ &$\ldots$\\
    +4E & \phantom{338}.5302 & 3.2     &$\ldots$ &$\ldots$\\
  $-$3E & \phantom{338}.5599 & 3.0     &$\ldots$ &$\ldots$\\
    +3E & \phantom{338}.5831 & 3.1     & 5.3     & 3.3    \\
    +1E & \phantom{338}.6150 & 3.3$^a$ & 7.0     & 4.4    \\ \hline
\multicolumn{5}{c}{$7_K-6_K$ band; A-type} \\ \hline
    +0A &           338.4086     & 4.8$^a$ & 7.5$^a$ & 4.4$^a$ \\
$\pm$4A/-2A & \phantom{338}.5127 & 3.3     & 7.7     & 5.4     \\
$\pm$3A & \phantom{338}.5419     & 4.6     & 7.7     & 5.6     \\
    +2A & \phantom{338}.6399     & 3.0     & 7.8     & 4.3    \\\hline
\end{tabular}\end{center}

``$\dagger$''The $5-4$ +3E and $-$2A lines are blended at 241.8430
GHz.  $^{a}$Lines asymmetric. No Gaussian fitted. Line width defined
as $\Delta v = \int T\,{\rm d}v\, /\, (1.064\, T_{\rm peak})$ where
the integral is over the total line profile and $T_{\rm peak}$ is the
temperature at the line peak.
\end{table}

\end{document}